%% file: motion.tex
\documentclass[journal]{IEEEtran}
\usepackage{blindtext}
\usepackage{graphicx}
\usepackage{mathtools}
\usepackage{amssymb}
\usepackage{afterpage}
\usepackage[colorlinks]{hyperref}
\usepackage[noadjust]{cite}
\usepackage{float}
\hypersetup{citecolor=blue,linkcolor=blue}
\usepackage[font=footnotesize]{caption}
\usepackage[labelformat=simple]{subcaption}
\usepackage{multirow}
\usepackage[ruled]{algorithm2e}
\usepackage{algpseudocode}

\graphicspath{{figs/}}
\input{macros}
\DeclareMathOperator*{\argmin}{arg\,min}
\DeclareMathOperator*{\argmax}{arg\,max}

\usepackage{amsmath}
\newcommand\norm[1]{\left\lVert#1\right\rVert}

\begin{document}
%
% paper title
% can use linebreaks \\ within to get better formatting as desired
\title{Motion Estimation Techniques for Volumetric Video Attribute Compression}
%
%
% author names and IEEE memberships
% note positions of commas and nonbreaking spaces ( ~ ) LaTeX will not break
% a structure at a ~ so this keeps an author's name from being broken across
% two lines.
% use \thanks{} to gain access to the first footnote area
% a separate \thanks must be used for each paragraph as LaTeX2e's \thanks
% was not built to handle multiple paragraphs
%

\author{ Haoran Hong, Eduardo Pavez, Antonio Ortega, Ryosuke Watanabe, Keisuke Nonaka
\thanks{Haoran Hong, Eduardo Pavez and Antonio Ortega are with the University of Southern California, 3740 McClintock Ave.,
Los Angeles, CA 90089-2564, USA (e-mail: haoranho@usc.edu; pavezcar@usc.edu; aortega@usc.edu).}
\thanks{Ryosuke Watanabe and Keisuke Nonaka are with KDDI Research, Inc., 2-1-15 Ohara, Fujimino, Saitama, 356-8502, Japan (e-mail: ru-watanabe@kddi.com; ki-nonaka@kddi.com).}
\thanks{}
\thanks{}
\thanks{}
}% <-this % stops a space

% note the % following the last \IEEEmembership and also \thanks - 
% these prevent an unwanted space from occurring between the last author name
% and the end of the author line. i.e., if you had this:
% 
% \author{....lastname \thanks{...} \thanks{...} }
%                     ^------------^------------^----Do not want these spaces!
%
% a space would be appended to the last name and could cause every name on that
% line to be shifted left slightly. This is one of those "LaTeX things". For
% instance, "\textbf{A} \textbf{B}" will typeset as "A B" not "AB". To get
% "AB" then you have to do: "\textbf{A}\textbf{B}"
% \thanks is no different in this regard, so shield the last } of each \thanks
% that ends a line with a % and do not let a space in before the next \thanks.
% Spaces after \IEEEmembership other than the last one are OK (and needed) as
% you are supposed to have spaces between the names. For what it is worth,
% this is a minor point as most people would not even notice if the said evil
% space somehow managed to creep in.

% The paper headers
\markboth{Journal of \LaTeX\ Class Files,~Vol.~14, No.~8, August~2021}%
{Shell \MakeLowercase{\textit{et al.}}: A Sample Article Using IEEEtran.cls for IEEE Journals}
%\IEEEpubid{0000--0000/00\$00.00~\copyright~2021 IEEE}

% The only time the second header will appear is for the odd-numbered pages
% after the title page when using the twoside option.
% 
% *** Note that you probably will NOT want to include the author's ***
% *** name in the headers of peer review papers.                   ***
% You can use \ifCLASSOPTIONpeerreview for conditional compilation here if
% you desire.

% If you want to put a publisher's ID mark on the page you can do it like
% this:
%\IEEEpubid{0000--0000/00\$00.00~\copyright~2007 IEEE}
% Remember, if you use this you must call \IEEEpubidadjcol in the second
% column for its text to clear the IEEEpubid mark.

% use for special paper notices
%\IEEEspecialpapernotice{(Invited Paper)}

% make the title area
\maketitle
\begin{abstract}
Point cloud compression relies on techniques to compress both geometry and attributes. 
Motion-based approaches for dynamic solid point cloud geometry compression within the geometry-based point cloud compression (G-PCC) framework have achieved significant reductions in geometry rate. 
However, motion-based techniques for attribute compression remain underexplored, making it challenging to achieve significant reductions in the temporal redundancy of attributes. 
Firstly, this paper proposes a geometry-based inter-coding scheme to compress the attributes of dynamic solid point clouds.
Secondly, a graph-based motion-estimation scheme for point-cloud attribute compression is proposed.
% , followed by a locally refined search.
Thirdly, an interpolation-free fractional-voxel motion estimation method is proposed to refine motion accuracy to fractional-voxel precision. 
Our experimental results on the MPEG point cloud dataset show that the proposed scheme outperforms G-PCC, GeS-TM, and V-PCC in lossless and lossy geometry conditions. 
We achieve average bitrate savings of $55.3\%$, $42.3\%$, and $16.5\%$ over G-PCC, GeS-TM, and V-PCC, respectively, under lossy-geometry conditions.
\end{abstract}


% For peer review papers, you can put extra information on the cover
% page as needed:
% \ifCLASSOPTIONpeerreview
% \begin{center} \bfseries EDICS Category: 3-BBND \end{center}
% \fi
%
% For peerreview papers, this IEEEtran command inserts a page break and
% creates the second title. It will be ignored for other modes.
\IEEEpeerreviewmaketitle

\section{Introduction}
\label{sec:introduction}
Recent progress in 3D acquisition and reconstruction technology enables the capture of 3D scenes, resulting in volumetric videos with high frame rates and resolutions. 
%Thus, volumetric videos, commonly represented by dynamic point clouds,  are an emerging media that has attracted much attention recently \cite{GROOT}. 
These volumetric videos have been applied across various fields, including 3D immersive telepresence, autonomous vehicle navigation, gaming, and animation \cite{Qian2021, overview2019}.
Point clouds, collections of points, each with its position (geometry) and associated attributes (e.g., YUV or RGB color), are well-suited for volumetric videos since they can represent arbitrary objects and scenes in three-dimensional space. Temporally acquired point clouds, arranged as an ordered set of frames according to timestamps, are called dynamic point clouds. 
When those points are dense enough to form continuous surfaces, dynamic point clouds are called dynamic solid point clouds \cite{solid}  (DSPCs). 
DSPCs in raw format would require a huge amount of bandwidth for transmission. Thus, there has been a significant interest in point cloud compression techniques, leading to two MPEG standardization efforts \cite{overview2019,overview2020}: 
Video-based point cloud compression (V-PCC) and geometry-based point cloud compression (G-PCC).

In V-PCC \cite{overview2020}, 3D DSPC are converted into 2D videos, which are then compressed using conventional video codecs, such as HEVC and VVC.
Key benefits of V-PCC include its reliance on existing 2D compression techniques to remove temporal redundancies among frames and its ability to incorporate future video coding innovations to further enhance performance.
However, 3D point clouds are usually a sampled representation of complex, non-developable surfaces whose geometry cannot be perfectly unfolded into 2D without distortion. 
Converting them into 2D atlases requires patch segmentation and projection that locally approximate the surface as piecewise planar regions.
These segmentation-and-projection operations can introduce point loss and temporal inconsistencies across frames, making lossless compression inefficient when those lost and inconsistent points must be preserved.
Normal-based segmentation cannot be consistent across frames because DSPCs undergo non-rigid deformation, self-occlusion, and viewpoint-dependent sampling changes, resulting in variations in surface normals and local topology over time.
As a result, regions that form a single patch in one frame may be split into multiple patches in the next, or vice versa.
This temporal inconsistency disrupts 3D voxel correlations and introduces artificial motion discontinuities, even when the underlying 3D motion is smooth.
The problem is further amplified by patch packing during projection, which 
arranges patches in a 2D atlas in a descending order of patch size. This may result in two spatially adjacent or motion-corresponding regions in 3D being placed at unrelated 2D locations in different frames.
Consequently, smooth 3D motion may appear to exhibit abrupt spatial jumps in 2D video.
Such discontinuities degrade inter-frame efficiency in HEVC/VVC, leading to unnecessary bitrate consumption due to both poorer prediction and lossless motion vector encoding of non-smooth motion fields. 
%. The situation is even worse in lossless compression.
% Besides, 3D-to-2D Projection can inevitably generate geometric distortion in order to facilitate coding performance.
% In summary, smooth 3D motion does not guarantee smooth 2D motion, which can potentially result in coding inefficiency.

In G-PCC, point clouds are directly encoded using 3D compression techniques, eliminating the need for a 3D-to-2D projection. 
%Since its attribute coding scheme does not rely on a particular geometry coding method, its performance is less sensitive to sparsity and uneven point distribution.
G-PCC was initially used for static point clouds and dynamically acquired LiDAR point clouds, leading to the development of several intra-coding approaches for attribute data \cite{raht,IRAHT2019,RAGFT,bytedance2023,Shashank2021,Shashank2022}. 
Following the release of the second version of 
G-PCC, an exploration \cite{sandri2023,GeSTM} to extend it for attribute compression in DSPCs was initiated. 
This led to the MPEG geometry-based solid test model (GeS-TM), in which block-based motion estimation (ME) and Region-Adaptive Hierarchical Transform (RAHT)-based motion compensation (MC) \cite{interRAHT2025} strategies are proposed for attribute inter-coding, resulting in significant bitrate savings compared to intra-only G-PCC. 
However, current G-PCC inter-coding techniques have major limitations. 
In some approaches \cite{Xu2020,SantosICIP2021}, inaccurate color compensation may occur because integer-precision motion is independently computed for each block based solely on geometry, i.e., without (i) utilizing color information, (ii) considering motion correlation across blocks, or (iii) allowing fractional motion displacements.
Second, systems that use both color and geometry for MC, such as GeS-TM using RAHT \cite{souto2023,bytedance_TF2023,sandri2023,interRAHT2025}, 
apply MC only in the top few levels of the octree hierarchy, i.e., on a coarsely approximated version of the point cloud, which makes it inherently less accurate than MC on the original full-resolution point cloud.

To address these limitations, we propose novel ME and MC tools for inter-coding. 
Our ME scheme achieves accurate fine-resolution MC by using a strategy that is essentially equivalent to ME after super-resolving the geometry and color.
In our approach, MVs for all blocks are jointly estimated in a progressive manner. 
First, we define a graph-based integer voxel ME (IvME), combining block-wise matching based \textit{on both color and geometry} with a \textit{graph-based motion smoothing} that constrains motion to be smooth across blocks. 
This is followed by \textit{a locally refined IvME} and \textit{interpolation-free fractional voxel motion estimation (IF-FvME)}. 
With our proposed FvME, fractional voxel displacements can be estimated without first generating (via interpolation) a higher resolution point cloud.
Our proposed ME/MC scheme can be combined with existing G-PCC transform methods.

Our work in this paper builds on our preliminary work \cite{IME2023,FME2022}, but introduces major improvements. 
In \cite{IME2023}, we proposed a block-based ME scheme in which the MV for each block is estimated independently. However, this independent estimation fails to capture spatial correlations and interactions among neighboring blocks. As a result, the estimated MVs lack local consistency, leading to visually and structurally undesirable artifacts in the motion-displaced point cloud, including tearing, cracking, and shape distortions, particularly at block boundaries. 
In this paper, we propose a graph-based motion estimation approach that estimates block motion globally by introducing a graph-based penalty term that promotes smooth, coherent MVs across neighboring blocks, thereby achieving more accurate motion compensation.

In \cite{FME2022}, we introduced an \textit{interpolation-based} FvME method, where fractional-voxel motion compensation is performed by explicitly interpolating reference samples and evaluating the FvMC error at discrete candidate fractional positions. 
Although effective, this approach requires explicit sub-voxel interpolation and exhaustive FvMC error evaluation for all candidate fractional displacements, which becomes expensive when finer fractional precision is desired.
In this paper, we propose an \textit{interpolation-free} (IF) FvME scheme that reformulates fractional motion estimation as a global optimization problem over interpolation coefficients. 
Starting from the integer motion field obtained by the proposed graph-based IvME, each block first collects 27 local integer-voxel MC predictors from the neighborhood around its IvMV. 
The fractional-voxel predictor is then generated as a weighted combination of these IvMC predictors, where the weights correspond to interpolation coefficients rather than explicitly interpolated sub-voxel samples. 
Instead of directly searching for fractional displacement for each block, IF-FvME optimizes the interpolation coefficients by minimizing an MC residual cost. 
% Specifically, the objective includes both the prediction residual energy and a graph-Laplacian regularization term on the residual signal, encouraging the resulting MC residuals to be not only small in magnitude but also spatially smooth across blocks and easier to code. 
% Therefore, the proposed IF-FvME directly targets coding-friendly residual generation through global optimization, rather than merely selecting the fractional position with the smallest block-wise prediction error.
This formulation provides fractional-precision MC without the computational overhead of explicit sub-voxel interpolation and exhaustive FvMC error evaluation. 
It also enables finer motion precision, e.g., $\frac{1}{4}$-voxel motion, with significantly lower complexity than \cite{FME2022}. 
The proposed IF-FvME can be viewed as a 3D point-cloud generalization of interpolation-free fractional motion estimation in conventional video coding, where the fractional motion vector is inferred from a local model of the MC error surface rather than by direct fractional interpolation \cite{HEVC2020}. 
Furthermore, unlike our previous approach \cite{FME2022}, which operated on a precomputed IvMV database generated by exhaustive search \cite{IMVdataset}, the proposed IF-FvME is integrated with our graph-based IvME scheme. 
Together, they form a unified integer-to-fractional motion estimation pipeline that jointly exploits color-geometry matching, global motion-field optimization, and coding-oriented residual regularization.

Unlike in \cite{IME2023,FME2022}, where we used transform coding based on the graph Fourier transform, our proposed IvME and IF-FvME schemes are combined with the G-PCC \cite{GPCC} transform coding framework and can operate with either lossless or lossy geometry coding. 
The differences between  \cite{FME2022,IME2023} and this work are summarized in \autoref{tab:comparison_with_previousworks}.
Our main contributions are:
\begin{itemize}
\item A voxel-domain attribute inter-prediction pipeline including fractional precision ME and MC. This pipeline is integrated into the latest G-PCC codec to encode attributes on either lossless or lossy coded geometry, demonstrating the potential of geometry-based approaches for point cloud attribute coding (\autoref{sec:pipeline}).
\item  
IvME: a graph-based motion estimation scheme for DSPC attribute compression that promotes motion smoothness across blocks (\autoref{sec:ME}).
\item  
FvME: an interpolation-free fractional-precision ME scheme for DSPC attribute compression that avoids explicit point-cloud super-resolution and optimizes block-wise FvMVs in a continuous interpolation coefficient domain to improve residual coding efficiency (\autoref{sec:FvME}).
\item An experimental setup that isolates the effect of lossy geometry coding in attribute compression by comparing the proposed methods under different types of geometry compression, thus providing a fair comparison with the state-of-the-art in attribute coding  (\autoref{sec:experiments}).
\end{itemize} We achieve significant gains over state-of-the-art V-PCC, G-PCC, and GeS-TM under lossy and lossless geometry coding conditions (\autoref{sec:experiments}), as well as our earlier work \cite{IME2023,FME2022}. 

%\section{Related inter-coding works in G-PCC}
\section{Related work}
DSPCs represent object surfaces that can be deformed and translated over time, so the number of occupied voxels and their distribution within a 3D volume can vary. 
These temporal changes and irregular spatial sampling complicate the establishment of a simple correspondence between voxels across consecutive frames, making the development of ME and MC techniques for G-PCC more challenging than for V-PCC. 
In particular, while conventional video codecs estimate motion by comparing color information from blocks of the same size (same number of pixels), an ME estimation algorithm for G-PCC must compare regions where voxels are sparse in 3D, and where their numbers and positions differ across consecutive frames due to non-rigid motion. Thus, to successfully achieve ME/MC for PCs, both color and geometry matching should be considered.
Next, we review existing ME and MC schemes for 3D point clouds.

\begin{table*}[t]
  \centering
\begin{tabular}{|c||c|c|c|}
\hline
Techniques &  IvME & FvME& Residue coding  \\
 \hline
\cite{FME2022} & Database motion \cite{IMVdataset} & Interpolation-based  & GFT\cite{GFT2014}+RLGR\cite{RLGR} \\
%  \hline
% Locally optimized IvMV & Yes & Yes  & Yes \\
% \hline
% Motion precision & 1/2 & Integer &  1/4 \\
\hline
\cite{IME2023} & Block based Color-ICP & Adaptive graph filtering  & GFT\cite{GFT2014}+RLGR\cite{RLGR} \\
\hline
Proposed & Graph-based isometric ME & Interpolation-free & G-PCC\cite{GPCC} \\
\hline
\end{tabular}
 \caption{Comparison of our proposed approach with our previous approaches in \cite{FME2022} and \cite{IME2023} } 
 \label{tab:comparison_with_previousworks}
 \vspace{-8mm}
\end{table*}

\subsection{Motion Estimation (ME)} 
\textit{Block matching based ME} methods for 3D point clouds  \cite{Ricardo2017,Camilo2018,Camilo2019} mimic video coding approaches by using 3D block-wise comparison. 
Each predicted block is exhaustively compared with all candidate blocks using a \textit{color and geometry distance criterion}. The block with the minimum distance is considered the best match, and the corresponding displacement is the estimated motion. 
% Since the complexity scales exponentially with the size of the search area, these schemes can be computationally prohibitive for large search ranges. 
Since the ME distance criterion depends on two distinct point cloud characteristics (color and geometry), optimization involves evaluating all candidate displacements via exhaustive search. This leads to high computational complexity, making such methods potentially prohibitive for large search ranges. Furthermore, these methods estimate motion independently for each block, which can lead to inconsistent motion vectors across neighboring blocks and blocking artifacts, thereby reducing coding efficiency. 
Our proposed approach overcomes the limitations of these methods by \textit{utilizing geometry and color matching in two separate stages} to reduce complexity, enable practical fractional ME, and produce spatially coherent motion across blocks.  

In \textit{point registration-based} ME, two types of registration techniques are considered. 
In \textit{graph matching} \cite{Dorina2016}, spectral matching is applied between the sampled voxels in the reference and predicted frames to establish voxel-wise correspondence. The motion for the remaining voxels is computed based on their positions relative to the globally matched sampled voxels, subject to a geometry-smoothness constraint. 
%However, point clouds are discrete samples of object surfaces, and sampled points generally differ across frames. Ground truth one-to-one matching does not always exist.
\textit{Geometric-ICP} applied to each prediction unit \cite{Rufael2017,Xu2020,SantosICIP2021} extracts the nearest point set and estimates rigid motion (e.g., rotation and translation) iteratively. 
For efficient signaling, the rotation matrix is represented as a quaternion. 
Note that both point registration techniques ignore color during ME, which may lead to suboptimal motion estimates for color prediction. 
Moreover, these approaches do not guarantee a smooth motion field. 
While these methods rely solely on geometry and estimate each block’s motion independently, our graph-based motion estimation scheme integrates a color-geometry similarity metric and globally optimizes block-wise motion field through a regularization term. 
This yields motion vectors that are not only better aligned with color attributes but are also spatially coherent, resulting in a more reliable initialization for subsequent color-driven refinement. 

Recently, \textit{learning-based} ME has also been explored for point cloud video compression.
U-Motion \cite{UMotion2024} proposes a learned point cloud video codec with a U-structured, multiscale inter-frame prediction module, in which explicit motion estimation and compensation are performed in the latent feature domain.
% Instead of searching for block-wise displacements or estimating rigid transforms, U-Motion extracts motion embeddings from the current and reference-frame latent features at multiple octree scales.
% Through top-down and bottom-up connections, motion information from both higher and lower scales is used to estimate the motion at the current scale.
% The decoded motion field is then used to warp the reference-frame latent features, and interpolation is applied to generate temporal context for predictive coding.
Although this learned framework improves temporal prediction by exploiting multiscale motion features, it relies on neural network training, latent-domain motion representations, and learned entropy models, which makes it less interpretable and less directly controllable than explicit optimization-based ME.
In addition, the estimated motion is implicitly optimized via rate-distortion training rather than by directly enforcing block-level color-geometry matching and spatial smoothness constraints.

% While existing block matching, point registration, and learning-based methods provide different mechanisms for exploiting temporal redundancy, they either suffer from high search complexity, ignore color information, estimate motion independently without sufficient spatial coherence, or rely on implicit learned representations.
% In contrast, our graph-based motion estimation scheme integrates a color-geometry similarity metric and globally optimizes the block-wise motion field. This produces motion vectors that are better aligned with color attributes while remaining spatially coherent, leading to more compact and easier-to-code prediction residuals.

\subsection{Motion Compensation (MC)}
%Two main approaches have been considered.  
%In the first class, MC is applied in the transform domain. 
\textit{Transform-domain MC} predicts the transform coefficients of the current frame from those of motion-compensated frames. Since RAHT is the most widely adopted transform in attribute coding, e.g., in G-PCC, several transform-domain MC schemes \cite{souto2023,sandri2023,bytedance_TF2023,interRAHT2025} have been developed based on RAHT.
However, due to geometric discrepancies and the exponentially increasing signaling overheads, RAHT-domain MC is generally restricted to the upper levels of the octree, 
%and cannot be extended to the lower or leaf-node level, 
which limits its benefits for coding because coarser compensation results in higher residual energy than finer compensation.  

Alternatively, in \textit{voxel-domain MC} methods, motion-displaced reference point attributes are mapped onto the current frame using nearest neighbor projections. In previous work \cite{IME2023,FME2022}, we show that gains over intra-coding are not guaranteed for all blocks undergoing inter-coding. Many blocks may suffer from degraded prediction quality, resulting in overall rate-distortion loss.
Previous work has addressed this issue by allowing per-block inter- and intra-mode decisions \cite{Souto2020}.
As an alternative, filtering approaches have been proposed to improve the quality of the predictors, by averaging attributes from multiple nearest neighbor voxels for MC \cite{Dorina2016} or  
applying the spatial-temporal Gaussian Markov random field (GMRF) model of  \cite{zhang2015graph} to the nearest neighbor predictors \cite{Xu2020}.
Both can be viewed as techniques using weighted averaging of multiple IvMCs.
However, the effectiveness of GMRF-based prediction strongly depends on how well the underlying color covariance structure conforms to the assumed GMRF, which is constructed solely from geometric information and may not accurately capture attribute variation.
In our previous work \cite{FME2022}, we proposed adaptive weighted averaging of multiple reference attributes through fractional voxel interpolation. While this improves prediction quality, it comes at the cost of significant computational complexity.
Each finer fractional precision increases the number of fractional voxels by a factor of eight, requiring repeated nearest-neighbor searches through kd-tree traversals. 

Our proposed interpolation-free FvME generalizes existing techniques for weighted averaging of IvMCs. It uses the fact that the unknown color of a fractional voxel is estimated through trilinear interpolation from its surrounding integer-voxel neighbors, where the interpolation weights depend on the fractional motion vector. Rather than performing voxel interpolation for each candidate fractional motion vector and selecting the best match, we directly estimate the optimal real-valued interpolation weights on a convex set, then compute the best fractional displacement that matches them.
This perspective results in lower energy residuals and a more computationally efficient FvME,  because no explicit interpolation or costly spatial searches are required.

\begin{figure}[tbh!]
\begin{center}
\includegraphics[width = 0.7\linewidth]{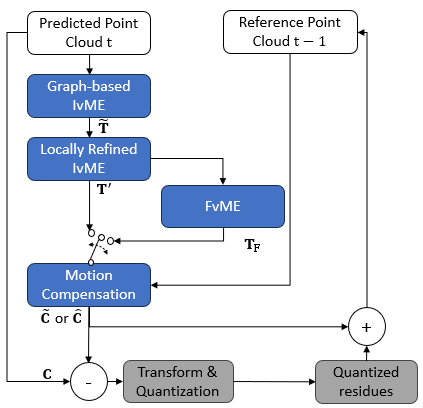}
\caption{The pipeline of the motion-based inter-coding scheme. The blue boxes represent the proposed inter-prediction techniques. Grey boxes represent functions identical to those in G-PCC.}
\label{fig:framework}
\end{center}
\end{figure}

\section{Pipeline overview}
\label{sec:pipeline}
We focus on 3D block-based ME/MC for attribute compression within the G-PCC framework.
As in G-PCC and V-PCC, the geometry is encoded first, followed by recoloring and color compression. 
Our proposed scheme is compatible with both lossless and lossy geometry coding and 
assumes that the decoded geometry is available at the encoder and the decoder.  
When geometry is encoded losslessly, the original point-to-color correspondence remains intact, and recoloring is unnecessary. 
However, under lossy geometry coding, both the number and spatial distribution of points may change, disrupting the original one-to-one mapping between points and their associated color attributes.
Thus, a recoloring step transfers color values from the original point cloud to the new point cloud obtained from reconstructed (coded) geometry. 
Next, we provide an overview of our approach (\autoref{fig:framework}).

\subsubsection{Block partition}
A dynamic point cloud is a sequence of $T$ point clouds, each of which is referred to as a frame.  
The frame at time $t$ has $N_t$ points, which can be represented as an unordered point set $P_t$.
Each point is $\pv \in P_t$ is denoted by its Cartesian coordinates $\pv = (x,y,z)$.
Its associated color attribute is denoted by $\cv(\pv)$ in YUV format.
The predicted (or current) frame $P_t$ is partitioned into non-overlapping 3D blocks (e.g., using an octree). The $b$-th block currently being predicted is denoted by the set $P_{b}\subseteq P_t$. The reference frame is the previously coded frame, denoted by $P_{t-1}$. 
For each block, we estimate an isometry-regularized motion, defined as a block-level translation that approximates isometry deformation. This motion provides a physically plausible alignment that closely approximates the underlying motion between $P_b$ and a corresponding region in $P_{t-1}$, and serves as the basis for subsequent motion estimation tailored for color prediction.

\subsubsection{Graph-based integer voxel motion estimation (\autoref{ssec:Graph_IvME})}  
The first stage is \textit{graph-based integer voxel motion estimation (IvME)}, which formulates motion estimation as a global optimization problem over all blocks.
Specifically, IvME performs a color-aware ICP search over a large region in the reference frame, while incorporating a graph-based regularization term to enforce structural consistency. The objective function consists of a fidelity term and a regularization term. The fidelity term measures geometric and color consistency between the current and reference frames, while the regularization promotes smooth motion across blocks and preserves structural consistency during fitting. 

\subsubsection{Locally-refined integer voxel motion estimation (\autoref{ssec:LR_IvME})} 
In the previous step, block motion is estimated over a large candidate region by jointly considering geometry, color, and graph-based smoothness. 
However, while jointly using geometry and color is beneficial for recovering a physically meaningful and realistic motion field, attribute coding ultimately benefits from minimizing the color prediction residual in the reconstructed reference.
Thus, to further improve graph-based motion in terms of color prediction, we propose a \textit{Locally Refined IvME} that starts from the estimate from the previous step and refines it within a small neighborhood so as to be locally optimal for color prediction, while largely preserving the geometric consistency and cross-block smoothness imposed by the previous step.   
Furthermore, this two-step strategy is computationally more efficient than exhaustively optimizing a color-only objective over the full search region, while still exploiting geometry, color, and cross-block smoothness to obtain a physically meaningful initialization \cite{Xu2020,Ricardo2017}.

\subsubsection{Integer voxel motion compensation (\autoref{ssec:ME_overview})}
After IvME, the predicted block is displaced by $\mv$ within the reconstructed reference frame. MC is then carried out by locating, for each voxel in the current frame, the nearest neighbor at the displaced position in the coded reference frame and using the attributes of the nearest motion-displaced neighbors to predict the current block, thereby yielding IvMC. Note that the geometry for $P_t$ and $P_{t-1}$ is available at the decoder and therefore given the motion $\mv$ for any point in a block in $P_t$, it is possible to search for the nearest points in $P_{t-1}$. 

\subsubsection{Fractional voxel motion estimation and compensation (\autoref{sec:FvME})}
FvME is an optional refinement tool designed to improve the resolution or precision of IvMVs from integer-voxel accuracy to fractional-voxel accuracy. 
Instead of explicitly generating sub-voxel point-cloud samples and evaluating the FvMC error at each candidate fractional position, the proposed interpolation-free FvME first estimates interpolation weights that form a convex combination of the IvMC predictors around each integer motion point. 
These weights are independently optimized within blocks by minimizing motion-compensation residuals.
Next, the estimated weights are quantized to the nearest feasible trilinear coefficient vector on a predefined fractional-resolution grid. Each grid point corresponds to a specific fractional displacement, i.e., a candidate FvMV. 
The quantized trilinear coefficients associated with the selected grid point are used to generate the final FvMC through the trilinear interpolation.
Note that the interpolation-free FvME also improves ME efficiency. An exhaustive search over all discrete fractional displacements may outperform our interpolation-free approach, but at a significantly higher cost. 

\subsubsection{Transform and Quantization}
After motion compensation, the residues are the difference between the original color signal and the motion-compensation color signal. These residual color attributes are then encoded using the standard coding tools in G-PCC \cite{GPCC}: attribute transform, e.g., RAHT, and entropy coding, e.g., run-length Golomb-Rice coder.  
%(\autoref{fig:framework}).

\section{Integer Voxel Motion Estimation and Compensation}
\label{sec:ME} 
Our proposed IvME algorithm consists of two steps.   
The first step, graph-based motion estimation, is summarized in \autoref{alg:IvME} and described in detail in  \autoref{ssec:Graph_IvME}. 
Graph-based IvME produces a motion vector per block 
$\tilde{\tv} \in \mathbb{R}^{3\times 1}$. 
These MVs are obtained from color and geometry information of both the predicted and reference point clouds and are regularized using graph-based distance preservation.  
In the second step, since the motion must be used for color compression, we 
introduce a local motion estimation refinement to obtain a new motion vector $\hat{\tv} \in \mathbb{R}^{3\times 1}$ to reduce the energy of the color residuals. 
For this, we first explain how to perform MC from the nearest neighbor at the motion-displaced position and obtain color residuals (\autoref{ssec:ME_overview}).
In the motion refinement step (\autoref{ssec:LR_IvME}), the IvME of each block $\hat{\tv}$ is refined to minimize the energy of the MC color residual, 
resulting in the overall integer motion for each block:
\begin{equation}
\label{equ:m}
 \tv' = \tilde{\tv} + \hat{\tv}.
\end{equation}
% We first introduce graph preliminaries \autoref{ssec:Graph_Preliminaries} and our graph-based motion estimation scheme \autoref{ssec:Graph_IvME}, which produces a coarse integer motion estimate $\hat{\tv}$.
% The attribute MC is introduced in \autoref{ssec:MC}, which is used for locally refined IvME introduced in \autoref{ssec:LR_IvME}.
% The coarse integer motion $\hat{\tv}$ is then locally refined to directly minimize the MC residual, leading to an additional local shift $\tilde{\tv}$. The overall integer voxel motion is:
% \begin{equation}
% \label{equ:m}
%  \mv = \hat{\tv} + \tilde{\tv}.
% \end{equation}

\subsection{Graph Representations for Point Clouds}
\label{ssec:Graph_Preliminaries}
We use graphs to connect spatially neighboring points in a point cloud, enabling us to apply graph signal processing for ME. %The graph structure provides local neighborhood relationships, enabling gradient and Laplace operators on the discretized surface. 
A graph is defined as $G=(V,E)$, where $V$ is the set of $N$ vertices and $E\subseteq V\times V$ is the set of edges. 
The $i$-th vertex is denoted  $v_i\in V$ and the edge connecting $v_i$ and $v_j$ is denoted as $e_{ij}\in E$. The weight of an edge is denoted as $w_{ij}$.
We use the $n \times n$ \textit{combinatorial Laplacian} matrix, $\Lm = \Dm-\Wm = \Um \Lambdam \Um^\transp$, where the \textit{degree matrix} $\Dm$ is a $n\times n$ diagonal matrix with $D_{i,i} = d_i = \sum_j w_{ij}$, and $\Um$ and $\Lambdam$ are the eigenvector and eigenvalue matrices of $\Lm$, respectively. 
Given a graph signal $\fv \in \mathbb{R}^n$, its smoothness is quantified by:
\begin{equation}
\label{equ:dirichilet_L}
\fv^\top\Lm\fv  =  \sum_{i,j: e_{ij}\in E} w_{ij}(\fv[i]-\fv[j])^2, 
% =  \frac{1}{2}\sum_{i\in V} \norm{\nabla_i \fv}^2.
\end{equation}
which is smaller (i.e., has lower variation) when $\fv$ is smoother. 

\subsection{Motion-based Inter Coding Overview}
\label{ssec:ME_overview}
In this subsection, we explain how to obtain the MC color residuals from arbitrary block-based motion vectors, such as the ones from \cite{IME2023,IMVdataset}. In the following subsections, we explain our proposed approach for obtaining these motion vectors. 
The voxel coordinates of the current  and reference frames are represented by the matrices
\begin{align}
\label{equ:P}
\Pm_t &= (\pv_{1}, \dots\pv_{N_t})^\top \in \mathbb{R}^{N_t \times 3},\\ %\\
% \tilde{\Pm} &= (\tilde{\pv}_{1}, \dots, \tilde{\pv}_{n})^\top.
\label{equ:P_ref}
\Pm_{t-1} &= (\pv'_{1}, \dots, \pv'_{N_{t-1}})^\top \in \mathbb{R}^{N_{t-1} \times 3}. %\\
% \tilde{\Pm} &= (\tilde{\pv}_{1}, \dots, \tilde{\pv}_{n})^\top.
\end{align}
We will also denote the voxels as  sets $P_t = \lbrace \pv_1,\cdots, \pv_{N_t}\rbrace$ and $P_{t-1} = \lbrace \pv'_1,\cdots, \pv'_{N_{t-1}}\rbrace$, respectively.
To simplify the notation, we replace $N_t$ by $N$ in what follows. 
To obtain the block structure, the voxelized points in $P_t$ are first sorted in Morton order so that spatially neighboring voxels within the same block are placed contiguously in the ordered sequence. 
An octree representation is then used to partition $P_t$ into $B$ non-overlapping blocks \cite{GFT2014,schnabel2006octree}. 
We index these blocks by $\{1,2,\dots,B\}$.
%Since $\hat{\Tm}$ is defined at the block level, 
The block assignment matrix $\Bm \in \{0,1\}^{N \times B}$  links each point $\pv_i$ in $P_t$ to its block: 
%, then for  $1\leq i \leq N_t$  and $1\leq b\leq B$, 
\begin{equation}
\label{equ:B}
\Bm_{ib} =
\begin{cases}
1, & \text{if } \pv_i \in P_b,\\
0, & \text{otherwise},
\end{cases}
\end{equation}
where $P_b \subset P_t$ is the  the set of points in the $b$th block.
Given a block-wise motion field $\Tm= (\tv_1,\;\cdots,\;\tv_{B})^\top$, the corresponding point-wise motion field $\Dm$ is:
\begin{equation}
\label{equ:D}
\Dm(\Tm) = \Bm\Tm = (\dv_1(\Tm),\;\cdots,\;\dv_{N}(\Tm))^\top \in \mathbb{R}^{N\times 3},
%=  \tilde{\Pm}_t - \Pm_t = (\tilde{\pv}_{1} - \pv_{1},\; \dots, \tilde{\pv}_n - \pv_n)^{\top}.
\end{equation}
where $\dv_{i}(\Tm) = \tv_b$ if $\pv_i \in P_b$.
The displaced coordinates of the current frame are expressed as:
\begin{equation}
\label{equ:tilde_P}
\tilde{\Pm}_t(\Tm) = \Pm_t + \Dm(\Tm) = (\tilde{\pv}_{1}(\Tm),\; \dots, \tilde{\pv}_{N}(\Tm))^{\top}.
\end{equation}

% With this definition, the displaced coordinates of the current frame are expressed as:
% \begin{equation}
% \label{equ:tilde_P}
% \tilde{\Pm}_t(\Tm) = \Pm_t + \Dm(\Tm) = (\tilde{\pv}_{1},\; \dots, \tilde{\pv}_{N})^{\top},
% \end{equation}
% where every $\pv_i \in P_b$ is displaced by the associated $\tv_b$.
% \subsection{Motion Compensation}
% \label{ssec:MC}
Note that, unlike 2D videos, whose pixel values lie on a regular 2D grid, 3D DPCs have irregular geometry, so only some positions in the regular 3D grid contain information.
Thus, for motion compensation of voxel $\pv_i$ in the current frame, given its point motion  $\dv_i$ ($i$th entry of $\Dm(\Tm)$), we use the nearest motion-displaced neighbor in the reference frame as $\pv_i(\Tm)$:
\begin{equation}
\begin{split}
\label{equ:NN}
\pv_i(\Tm) = \argmin_{\pv'\in P_{t-1}} \Vert \tilde{\pv}_{i}(\Tm) -  \pv' \Vert_2.
\end{split}
\end{equation}
The  coordinates of the motion-displaced nearest points in the reference frame are given by $\Qm$:
\begin{equation}
\Qm(\Tm) = (\pv_1(\Tm),\;\cdots,\;\pv_{N}(\Tm))^\top \in \mathbb{R}^{N\times 3}.
\end{equation}
Then, to predict the color information in $\pv_i$ we use the color of $\pv_i(\Tm)$ from  \eqref{equ:NN}: 
%is the color MC of $\pv_i$ denoted as $\tilde{\cv}_i(\Tm)$ which is also a function of block motion field $\Tm$:
\begin{equation}
\label{equ:MC}
 \tilde{\cv}_i(\Tm) = \cv(\pv_i(\Tm))\textnormal{, where }\pv_i(\Tm)\in P_{t-1}.
\end{equation}
Since the block motion $\Tm$ is transmitted as overhead and the geometries of frames $t$ and $t-1$ are available,  the decoder can use \eqref{equ:NN} to find $\pv_i(\Tm)$ for any point $\pv_i$ in frame $t$. 
The target colors of the current frame are stacked into the matrix:
\begin{equation}
\label{equ:C_target}
\Cm 
=\left(\cv(\pv_1),\;\cdots,\;\cv(\pv_{N})\right)^\top \in \mathbb{R}^{N\times 3},
\end{equation}
whose $i$-th row is the color of point $\pv_i$.
Likewise, stacking the MC of each point into matrix form, we get:
\begin{equation}
\label{equ:IvMC}
\tilde{\Cm}(\Tm) 
= [\tilde{\cv}_1(\Tm),\hdots,\tilde{\cv}_N(\Tm)]^\top  \in \mathbb{R}^{N\times 3},
\end{equation}
which is used to compensate $\Cm$ \eqref{equ:C_target} and compute the color residue, e.g., $\Cm - \tilde{\Cm}(\Tm)$, as in \autoref{fig:framework}.

This paper applies the  computation of color residuals explained above using the block-wise motion field given by:
\begin{equation}
\begin{split}\label{equ:T_prime}
\Tm'=(\tv'_1,\;\cdots,\;\tv'_{B})^\top = \tilde{\Tm}+\hat{\Tm} \in \mathbb{R}^{B\times 3}, 
\end{split}
\end{equation}
using the graph-based block-wise motion field (\autoref{ssec:Graph_IvME})  
\begin{align}
\label{equ:T_hat}
\tilde{\Tm} = (\tilde{\tv}_1,\;\cdots,\;\tilde{\tv}_{B})^\top \in \mathbb{R}^{B\times 3}, 
\end{align}
and the  locally refined block-wise motion field (\autoref{ssec:LR_IvME})  
\begin{align}
\label{equ:T_tilde}
\hat{\Tm} = (\hat{\tv}_1,\;\cdots,\;\hat{\tv}_{B})^\top \in \mathbb{R}^{B\times 3}.
\end{align}

\subsection{Graph-based Integer voxel Motion Estimation}
\label{ssec:Graph_IvME}
IvME aims to optimize ${\Tm'}$ in \eqref{equ:T_prime} to achieve the best color prediction between frames $\Pm_t$ and $\Pm_{t-1}$ . First, we propose an iterative procedure to find $\tilde{\Tm}$, starting with $\tilde{\Tm}_0 = \textbf{0}$, and finding  $\tilde{\Tm}_k$ for $k\geq 1$ until convergence (\autoref{alg:IvME}):
\begin{itemize}
\item \textbf{Step 1}: The block-wise motion $\tilde{\Tm}_{k-1}$ is used to deform the point cloud $\Pm_t$ leading to $\tilde{\Pm}_t(\tilde{\Tm}_{k-1})$ via \eqref{equ:tilde_P}.

\item \textbf{Step 2}: A many-to-one nearest neighbor matching is performed between $\tilde{\Pm}_t(\tilde{\Tm}_{k-1})$ and $\Pm_{t-1}$, which leads to the reference point-wise motion field $\tilde{\Dm}(\tilde{\Tm}_{k-1})$ (see \autoref{sssec:D}).

\item \textbf{Step 3}: The point-wise motion field $\tilde{\Dm}(\tilde{\Tm}_{k-1})$ is converted into a block-wise motion field, $\tilde{\Tm}_k$, by solving the optimization:
\begin{equation}
\label{equ:rough_model}
\begin{split}
\tilde{\Tm}_k =  \argmin_{\Tm} \left \| \Dm(\Tm) - \tilde{\Dm}(\tilde{\Tm}_{k-1}) \right \|^2_F + \beta,\text{reg}(\Tm,\Rm),
% &= \left \| \Tm - \hat{\Tm} \right \|^2_F + .
\end{split}
\end{equation}
where $\beta$ controls the tradeoff between two terms. 
The first term encourages the point-wise motion derived from block-wise vectors $\Dm(\tilde{\Tm}_k)$ to match the ``noisy'' unconstrained point-wise motion from nearest-neighbor matching $\tilde{\Dm}(\tilde{\Tm}_{k-1})$. The second term is a graph-based regularizer that promotes spatial smoothness and physically plausible deformation in ${\Dm}(\tilde{\Tm}_k)$. The auxiliary variable $\Rm$, to be defined below, helps offset the penalty of rotation deformation.  
% use k to represent iteration
%The obtained $\hat{\Tm}_k$ is used to update $\tilde{\Pm}_t$ in \textbf{Step 1} for next iteration. 
\end{itemize}
\begin{algorithm}
\caption{Graph-based IvME Algorithm}\label{alg:IvME}
\KwData{$P_{t-1}$ and $P_{t}$}
\KwResult{$\tilde{\Tm}=(\tilde{\tv}_1,\;\cdots,\;\tilde{\tv}_B)^\top$}
Initialize $\tilde{\Tm}_{0} = \textbf{0}$\;
\While{not converged or $k\leq k_{max}$}{
 \textbf{Step 1}: Calculate the motion-displaced coordinates of the current frame $\tilde{\Pm}_t$ via $\tilde{\Tm}_{k-1}$ and \eqref{equ:tilde_P}\;
 \textbf{Step 2}: Obtain reference dense motion field $\tilde{\Dm}(\tilde{\Tm}_{k-1})$ (see \autoref{sssec:D})\;
 \textbf{Step 3}: Optimize the block motion field in the current $k$-th iteration $\tilde{\Tm}_k$ via \eqref{equ:rough_model}\;
}
\end{algorithm}

The above three steps are repeated until convergence to produce the graph-based IvME  (\autoref{alg:IvME}). 
In the following subsections, we present the details of the generation of the reference motion field $\tilde{\Dm}$ (Step 2) and the formulation of the objective function \eqref{equ:rough_model} and its solution (Step 3).

\subsubsection{Reference Point-wise Motion Field}
\label{sssec:D}
%Point clouds are discrete samples of object surfaces, and sampled points generally differ across frames. Ground truth one-to-one matching does not always exist, so an estimated one-to-one correspondence between points is not completely reliable. Thus, we first perform a rough many-to-one nearest neighbor matching. 
Since our goal is color compression, we adopt a hybrid distance metric that jointly considers geometry and color.   
The hybrid distance between two points $\pv_i$ and $\pv_j$, given $\beta_p \in [0,1]$, is:
\begin{equation}
\label{equ:NNcriterion}
\begin{split}
\delta(\pv_i,\pv_j)&= \beta_p\Vert \pv_i - \pv_j\Vert^2 + (1-\beta_p)\Vert \cv(\pv_i) - \cv(\pv_j)\Vert^2. 
\end{split}
\end{equation}
Define the matrix $\Sm \in \mathbb{R}^{N_t \times N_{t-1}}$, where each entry is the distance between a displaced point, $\tilde{\pv}_{i}(\tilde{\Tm}_{k-1})$ from \eqref{equ:tilde_P}, in the current frame and a candidate point, $\pv'_j$ from \eqref{equ:P_ref}, in the reference frame\footnote{We use a kd-tree to find the nearest neighbors for $N$ points, which leads to a $N\times 1$ vector of the nearest point indices. The space complexity is $O(N)$ and time complexity is $O( N \log(N) )$, instead of $O(N^2)$}:
\begin{equation}
\Sm(\tilde{\Tm}_{k-1})_{ij} =\delta(\tilde{\pv}_{i}(\tilde{\Tm}_{k-1}),\pv'_j), \quad \tilde{\pv}_{i}\in \tilde{P}_{t},\; \pv'_j\in P_{t-1},
\end{equation}
The set of many-to-one matching matrices is:
\begin{equation}
\mathcal{P} = \left\{ \Xm \in \{0,1\}^{N_{t-1}\times N_t} \ \middle|\ \sum_{i=1}^{N_{t-1}} \Xm_{ij} = 1 \;\; \forall j \right\}. 
\end{equation}
The column-sum condition of the above ensures that every predicted point is assigned to exactly one reference point. Since the row sums are left free, multiple predicted points are allowed to match the same reference point, resulting in a many-to-one matching when solving:
\begin{equation}
\Mm(\tilde{\Tm}_{k-1}) = \arg\min\limits_{\Xm \in \mathcal{P}} \operatorname{Tr}(\Sm(\tilde{\Tm}_{k-1}) \Xm).
\label{equ:matching}
\end{equation}
Equivalently, this optimization independently selects the minimum composed distance \eqref{equ:NNcriterion}  for each predicted point.

The  coordinates of the matched points in the reference frame are given by $\tilde{\Qm}$, which is a function of $\tilde{\Tm}_{k-1}$:
\begin{equation}
\label{equ:Q}
\tilde{\Qm}(\tilde{\Tm}_{k-1}) = {\Mm(\tilde{\Tm}_{k-1})}^\top \Pm_{t-1},
\end{equation}
allowing us to obtain the reference displacement field $\tilde{\Dm}$ as:
\begin{equation}
\label{equ:D_ref}
\tilde{\Dm}(\tilde{\Tm}_{k-1}) =  \tilde{\Qm}(\tilde{\Tm}_{k-1}) - \Pm_t.
\end{equation}

\subsubsection{Graph-based Regularization}
\label{sssec:regularization}
In this section, we first express the graph Laplacian smoothness of the point-wise motion field as a pairwise local consistency measure. 
This pairwise measure is aggregated into a block-wise local-consistency form, which is then converted into a local distance-consistency form by offsetting the relative edge rotations between neighboring blocks.

The regularization term in \eqref{equ:rough_model} is designed to favor isometric deformation, that is, the geodesic distances within the original shape, given by $\Pm$, and the new motion-compensated shape, given by $\tilde{\Pm}$, are preserved. 
Since point clouds are discrete surface samples without explicit surface connectivity, exact geodesic distances are not directly available. Therefore, geodesic distance preservation is approximated via local Euclidean distance preservation, modeled using a sparse, locally connected graph.

For this, we build a distance-based graph on the point cloud $P_t$ with edge weights defined as: 
\begin{equation}
\label{equ:W}
w_{ij} =
\begin{cases}
1, & \text{if } \norm{\pv_i - \pv_j}_2 \leq \tau, \\
0, & \text{otherwise}.
\end{cases}
\end{equation}
We choose $\tau=1$, rather than $\tau= \sqrt{2}$ or $\tau = \sqrt{3}$. For voxelized point clouds, this connects each occupied voxel only to its face-adjacent neighbors along the principal axes of the canonical basis. 
This design ensures that the block-wise connectivity used later in this section remains sparse while preserving only the most relevant local interactions between directly adjacent blocks that share a face.

The smoothness of the point displacement field in \eqref{equ:D} is quantified using the   Laplacian quadratic form \eqref{equ:dirichilet_L}. From \eqref{equ:tilde_P} this can be expressed as the difference in point coordinates before ($\pv_i$) and after ($\tilde{\pv}_i$) deformation:
\begin{equation}
\label{equ:reg_displacement}
\begin{split}
\operatorname{Tr}(\Dm^\top\Lm\Dm) = \sum_{e_{ij}\in E} \norm{(\tilde{\pv}_i-\pv_i)-(\tilde{\pv}_j-\pv_j)}^2, 
\end{split}
\end{equation}
where we used the fact that $w_{ij}=1$ if $e_{ij}\in E$. 
%The regularization above penalizes large variations in the displacement vectors between neighboring points, thereby encouraging a smooth displacement field. 
Rearranging terms, we have:
\begin{equation}
\label{equ:reg_geodesic}
\begin{split}
\operatorname{Tr}(\Dm^\top\Lm\Dm) = \sum_{e_{ij}\in E} \norm{(\tilde{\pv}_i-\tilde{\pv}_j)-(\pv_i-\pv_j)}^2. 
\end{split}
\end{equation} 
This shows that the regularizer 
%penalizes changes in local edge vectors before and after deformation, 
encourages neighboring points to preserve their relative positions, providing a local-consistency prior for the estimated motion field.

Since we use block-wise motion, all points within a block share the same displacement and for any intra-block edge $(i,j)$, $\tilde{\pv}_i-\tilde{\pv}_j = \pv_i-\pv_j$. 
Thus, the metric in \eqref{equ:reg_geodesic} takes nonzero values for inter-block edges only.
Given a pair of blocks $(a, b)$, with point sets $P_a$ and $P_b$, we define the set of edges connecting them as:
\begin{equation}
\label{equ:inter_edges}
E_{ab} = \{(i,j)\mid (i,j)\in E, \pv_i\in P_a, \pv_j\in P_b, a \neq b \}, 
\end{equation}
allowing us to define the set of all inter-block edges as:
\begin{equation}
\label{equ:inter_edges2}
E_{\text{inter}} =\bigcup_{1 \leq a < b \leq B} E_{ab}.
\end{equation}
Thus, \eqref{equ:reg_geodesic} can be simplified to:
\begin{equation}
\label{equ:reg_geodesic_inter}
\begin{split}
\operatorname{Tr}(\Dm^\top\Lm\Dm) & = \operatorname{Tr}(\Dm^\top\Lm_{\text{inter}}\Dm)\\
&= 
\sum_{e_{ij}\in E_{\text{inter}}} \norm{(\tilde{\pv}_i-\tilde{\pv}_j)-(\pv_i-\pv_j)}^2,
\end{split}
\end{equation}
where $\Lm_{\text{inter}}$ is the Laplacian of the graph of inter-block edges. 

Note that different translations assigned to two neighboring blocks can change the orientation of their inter-block edges. In this case, the deformed edge vector $\tilde{\pv}_i-\tilde{\pv}_j$ may not be parallel to the original edge vector $\pv_i-\pv_j$, even when their lengths are the same. Such orientation changes can be unnecessarily penalized by \eqref{equ:reg_geodesic_inter}, which measures the norm of the difference between edge vectors rather than the difference in their lengths.

To eliminate this penalty, we compute a local rotation matrix $\Rm_{ab}$ that represents the optimal rotation from original to deformed edges for edges connecting blocks $a$ and $b$. Next, a rotation matrix $\Rm = (\Rm_1,\; \cdots,\; \Rm_{N_{\text{inter}}})^\top \in \mathbb{R}^{3N_{\text{inter}}\times 3}$ is formed by concatenating all inter-block edge rotation matrices, where $N_{\text{inter}}$ is the number of block pairs $(a,b)$ for which $E_{ab}$ is not empty.
This leads to our final choice of regularization for \eqref{equ:rough_model}, where, to avoid the rotation penalty of \eqref{equ:reg_geodesic_inter}, we use:
\begin{equation}
\begin{split}
\label{equ:reg_arap}
\text{reg}(\Tm,\Rm) = \sum_{a=1}^B\sum_{b=a+1}^B \sum_{e_{ij}\in E_{ab}} \norm{(\tilde{\pv}_i-\tilde{\pv}_j)-\Rm_{ab}(\pv_i-\pv_j)}^2 .
\end{split}
\end{equation}
By compensating for the local orientation change across each block boundary, $\Rm_{ab}$ rotates $\pv_i-\pv_j$ to be as close as parallel to $\tilde{\pv}_i-\tilde{\pv}_j$ as possible. \eqref{equ:reg_arap} reduces the effect of rotation and mainly penalizes non-isometric motion that stretches and compresses inter-block edge length \cite{ARAP2007}. Thus, the inter-block edges serve as elastic connections that promote isometry-preserving, coherent block motion during fitting.

\subsubsection{Explicit Form of the Objective Function \eqref{equ:rough_model}}
\label{sssec:objective}
Combining \eqref{equ:D} and \eqref{equ:D_ref}, the first term in \eqref{equ:rough_model} can be expressed as:
\begin{equation}
\begin{split}
\label{equ:D_fit}
 \left \| \Dm(\Tm) - \tilde{\Dm} \right \|^2_F &= \left \| \Bm\Tm - \tilde{\Dm} \right \|^2_F \\
 &= \operatorname{Tr}(\Tm^\top\Bm^\top\Bm \Tm 
- 2 \tilde{\Dm}^\top\Bm \Tm + \tilde{\Dm}^\top\tilde{\Dm}).
\end{split}
\end{equation}

We then define a sparse matrix $\Km\in\mathbb{R}^{N\times 3N_{\mathrm{inter}}}$ containing all the edge vectors.  
For each edge $(i,j)\in E_{ab}$, letting $m=m(a,b) \leq N_{\text{inter}}$ be the index of the corresponding block pair, the corresponding entries of $\Km$ are:
\begin{equation}
\begin{split}
\Km_{i,3m-2:3m} &\mathrel{+}= (\pv_i-\pv_j)^\top,\\
\Km_{j,3m-2:3m} &\mathrel{+}= -(\pv_i-\pv_j)^\top ,
\end{split}
\end{equation}
where $\mathrel{+=}$ denotes accumulation. If multiple inter-block edges associated with the same block-pair index $m$ are incident to the same point, their edge-vector contributions are summed into the same row segment of $\Km$.
Then $\Km\Rm\in\mathbb{R}^{N\times 3}$ contains the locally rotated original inter-block edge vectors for all points.
From \eqref{equ:reg_arap}, the second term in \eqref{equ:rough_model} can be written as:
\begin{equation}
\label{equ:reg_arap_matrix}
\begin{split}
&\text{reg}(\Tm,\Rm) = 2\operatorname{Tr}(\Tm^{\top}\Bm^{\top}\Lm_{\text{inter}}\Bm\Tm) \\
&+2\operatorname{Tr}(\Tm^{\top}\Bm^{\top}
(2\Lm_{\text{inter}}\Pm_t - \Km\Rm )) -2\operatorname{Tr}(\Pm_t^{\top}\Km\Rm).
% 2\operatorname{Tr}((\Pm + \Bm\Tm)^\top\Lm (\Pm + \Bm\Tm))   - 2\operatorname{Tr}\left ((\Pm + \Bm\Tm)^\top\Km\Rm \right ).
\end{split}
\end{equation}
%where $\Km\in \mathbb{R}^{N\times 3N_{\text{inter}}}$ is a matrix whose multiplication with $\Rm$ offsets the influence of the rotations on the deformed inter-block edges.

After combining \eqref{equ:D_fit} and \eqref{equ:reg_arap_matrix} and removing the constant term that is not related to $\Tm$ and $\Rm$,  \eqref{equ:rough_model} is equivalent to solving  the matrix optimization problem:
\begin{equation}
\begin{split}
\min_{\Tm,\Rm}&\operatorname{Tr}\!\left(\Tm^\top \Bm^\top(\textbf{I}+2\beta\Lm_{\text{inter}})\Bm \Tm\right) - 2\operatorname{Tr}(\beta\mathbf{P}^\top\mathbf{K}\mathbf{R}) \\
&+2\operatorname{Tr}\!\left(\Tm^\top\Bm^\top\left( 2\beta\Lm_{\text{inter}}\Pm_t - \beta\Km\Rm-\tilde{\Dm}\right)\right).
\label{equ:energy_model}
\end{split}
\end{equation}

\subsubsection{Optimization}
\label{sssec:optimization}
To minimize the objective function \eqref{equ:energy_model}, we adopt the alternating block coordinate descent that alternates between finding the optimal rotations $\tilde{\Rm}$ and finding the optimal block-wise motion field $\tilde{\Tm}$. 
We define the inter-block edge covariance matrix of blocks $a$ and $b$ as:
%$\Om_{ab}$
\begin{equation}
\Omegam_{ab} = \sum_{e_{ij}\in E_{ab}} (\pv_i-\pv_j)(\tilde{\pv}_i-\tilde{\pv}_j)^\top= \Um_{ab} \Sigmam_{ab} \Vm_{ab}^\top,
\end{equation}
where $\Um_{ab} \Sigmam_{ab} \Vm_{ab}^\top$ is the SVD of $\Omegam_{ab}$. 

For fixed $\Tm=\tilde{\Tm}$, minimizing \eqref{equ:energy_model} with respect to $\Rm$ is a linear optimization problem. The optimal solution of $\Rm$ in each iteration can be written in closed form as:
\begin{equation}
\label{equ:R}
\begin{split}
\tilde{\Rm}_{ab} &= \argmax_{\Rm_{ab}\in SO(3)}\langle \Rm_{ab},\Omegam_{ab} \rangle \\
&=\Um_{ab}
\begin{pmatrix}
1 & 0 & 0 \\
0 & 1 & 0 \\
0 & 0 & \det(\Um_{ab}\Vm_{ab}^\top) \\
\end{pmatrix}\Vm_{ab}^\top.
% \text{tr}\left (\hat{\Pm}^\top\Km\Rm \right ),
\end{split}
\end{equation}
where $SO(3)$ denotes the rotation group in $\mathbb{R}^3$. 

For fixed $\Rm=\tilde{\Rm}$, \eqref{equ:energy_model} turns into a quadratic optimization problem with respect to $\Tm$:
% \begin{equation}
% \label{equ:optimization_P}
% \hat{\Pm} = \argmin_{\tilde{\Pm}}\operatorname{Tr}(\tilde{\Pm}^\top (\textbf{I}+2\beta\Lm) \tilde{\Pm})   - 2\operatorname{Tr}\left (\tilde{\Pm}^\top(\beta\Km\hat{\Rm}+\Qm )\right ).
% \end{equation}
\begin{equation}
\label{equ:optimization_T}
\begin{split}
\tilde{\Tm} 
= &\argmin_{\Tm} \; 
 \operatorname{Tr}\!\left(\Tm^\top \Bm^\top(\textbf{I}+2\beta\Lm_{\text{inter}})\Bm \Tm\right) \\
&+2\operatorname{Tr}\!\left(\Tm^\top\Bm^\top\left( 2\beta\Lm_{\text{inter}}\Pm_t - \beta\Km\Rm-\tilde{\Dm}\right)\right), 
\end{split}
\end{equation}
so the optimal motion field can be written in closed form as:
\begin{equation}
\label{equ:T}
\begin{split}
\tilde{\Tm}=\left(\Bm^\top\left(\textbf{I}+2\beta\Lm_{\text{inter}}\right)\Bm\right)^{-1}\Bm^\top( \tilde{\Dm} + \beta(\Km\tilde{\Rm}-2\Lm_{\text{inter}} \Pm_t)).
\end{split}
\end{equation}

To further improve motion-compensated shape fitting in the reference frame, we can alternately update $\tilde{\Tm}$ and $\tilde{\Rm}$ for $l_{\max}$ inner iterations within each outer iteration $k$. Specifically, at the beginning of the $k$-th iteration, we initialize $\tilde{\Rm}_{k,0} = \tilde{\Rm}(\tilde{\Tm}_{k-1})$ via \eqref{equ:R} and then $\tilde{\Tm}_{k,0} = \tilde{\Tm}(\tilde{\Rm}_{k,0})$ via \eqref{equ:T}. For the $l$-th inner iteration, where $0 < l \leq l_{\max}$, we first update $\tilde{\Rm}_{k,l} = \tilde{\Rm}(\tilde{\Tm}_{k,l-1})$ via \eqref{equ:R}, and then update $\tilde{\Tm}_{k,l} = \tilde{\Tm}(\tilde{\Rm}_{k,l})$ via \eqref{equ:T}. After $l_{\max}$ inner iterations, the final estimates for the $k$-th iteration are given by $\tilde{\Rm}_{k} = \tilde{\Rm}_{k,l_{max}}= \tilde{\Rm}(\tilde{\Tm}_{k,l_{max}-1})$ and $\tilde{\Tm}_{k} = \tilde{\Tm}_{k,l_{max}} = \tilde{\Tm}(\tilde{\Rm}_{k})$.

\subsection{Locally Refined Integer voxel Motion Estimation}
\label{ssec:LR_IvME}
%The graph-based IvME stage estimates motion over a large candidate region by jointly considering geometry, color, and graph regularization. 
The previous IvME stage is designed to recover a geometrically meaningful and realistic motion field, rather than to directly minimize the final color-prediction residual on the reconstructed reference. 
%, particularly when that reference is affected by lossy compression.
To reduce the energy in this residual, 
%we improve the globally optimized motion by finding the refinement motion vectors $\tilde{\Tm}$ defined in \eqref{equ:T_prime}) using reconstructed colors to further minimize the MC error of \eqref{equ:MC}.  
for each block $b$, we take \eqref{equ:T} as the initial motion $\hat{\tv}_b$ from the previous stage and apply an exhaustive search within a small neighborhood to adapt it to the color-prediction objective.
Each coordinate of a locally refined motion vector is restricted to be an integer in $\left[ -1, 1 \right]$. Therefore, the candidate set for the local refinement of a \emph{single block} is
\begin{equation}
\label{equ:M}
\begin{split}
    \mathcal{M} &= \left\{ \tv = (t_x, t_y, t_z) \in \mathbb{Z}^3 \;\middle|\; -1 \leq t_x, t_y, t_z \leq 1 \right\}\\
    &=\left\{\tv_{(1)},\ldots,\tv_{(27)}\right\},
\end{split}
\end{equation}
which contains $27$ candidate MVs and $\tv_{(1)},\ldots,\tv_{(27)}$ are a fixed enumeration of the $27$ integer offsets.
For block $b$, the refined local offset $\tilde{\tv}_b$ is selected from $\mathcal{M}$. Stacking the selected offsets of all $B$ blocks yields the refined block-wise motion field $\tilde{\Tm}$, i.e., $\tilde{\Tm} = (\tilde{\tv}_1, \ldots, \tilde{\tv}_B)^\top$ with $\tilde{\tv}_b \in \mathcal{M}$ for all $b$. Equivalently, $\tilde{\Tm}$ is searched over
\begin{equation*}
    \mathcal{M}_T = \left\{ \Tm = (\tv_1, \ldots, \tv_B)^\top \;\middle|\; \tv_b \in \mathcal{M},\ \forall b \right\}.
\end{equation*}
The refined motion field is then chosen to minimize the color prediction error of the MC predictor in \eqref{equ:MC}:
\begin{equation}
\label{equ:m1}
    \hat{\Tm} = \arg\min_{\Tm \in \mathcal{M}_T} \left\| \Cm - \tilde{\Cm}(\tilde{\Tm}+\Tm) \right\|_F^2.
\end{equation}

% \begin{figure}[ht]
% \begin{center}
% \includegraphics[width = 0.3\textwidth]{./figures/ME2.png}
% \end{center}
% %\vspace{-4mm}
% \caption{Proposed IvME scheme.}
% \label{fig:ME}
% %\vspace{-5mm}
% \end{figure}

\section{Fractional Voxel Motion Estimation and Compensation}
\label{sec:FvME}
The concept of ME with fractional-precision MVs, widely used in video coding \cite{HEVC2020}, can be extended to point clouds. When an MV has sub-voxel components, the motion-compensated predictor must be evaluated at fractional locations, where no reference voxels exist, and the attribute values at these locations are obtained as linear combinations of the attributes of nearby integer voxels. In conventional video, interpolation-free fractional ME exploits the regular pixel grid. The fractional displacement is inferred directly on the image plane from a local model of the MC error surface. Point clouds lack such a regular structure, so this approach does not apply directly. Instead, our method estimates the fractional displacement in the interpolation-coefficient domain.
 
Weighted combinations of integer-voxel predictors have already proven effective for point cloud MC, e.g., by averaging the attributes of multiple nearest neighbors \cite{Dorina2016} or by applying model-based filtering to nearest-neighbor predictors \cite{Xu2020}. 
Since all of these methods compute weighted averages of multiple IvMC predictors, they can be interpreted as implicit forms of fractional-voxel MC. However, their averaging weights are either fixed \cite{Dorina2016} in advance or derived from a predefined graph model constructed solely from geometry \cite{Xu2020}. 
The FvME proposed in this section generalizes these approaches by directly optimizing averaging weight coefficients for each block. 
The philosophy is that, since the FvMVs are explicitly optimized for each block and each FvMV corresponds to a unique set of combination weights, FvMC can be viewed as an adaptive averaging operation tuned to the local attribute statistics.

The proposed FvME can also be viewed as an interpolation-free scheme. In our previous interpolation-based FvME \cite{FME2022}, \textit{for each candidate fractional motion displacement}, voxels at fractional positions were first explicitly generated by interpolating the geometry and color of neighboring integer voxels. 
Then, the MC error was evaluated at each candidate fractional displacement using interpolated values, so complexity increased with the number of candidate fractional motions. 
In contrast, the proposed method does not generate voxels at all fractional-displacement positions. Instead, we express the FvMC predictor as a trilinear combination of IvMC predictors (\autoref{ssec:FvME_formulation}), optimize the combination weights over a convex relaxation of the feasible set, and map the optimized weights back to trilinear weights and therefore a valid FvMV (\autoref{ssec:FvME_opt}). FvMCs are only computed after this final step, resulting in a significant reduction in computational complexity.

\subsection{FvMC modeling}
\label{ssec:FvME_formulation}
The integer MV computed in \autoref{ssec:LR_IvME} for  block $b$ is denoted by $\tv'_b$ (see  \eqref{equ:T_prime}).
%, which displaces the block to its integer motion position in the reference frame. 
 \autoref{fig:integer_8cubes}a depicts the integer vector $\tv'_b$ (in red) surrounded by eight unit cubes whose vertices lie on the integer grid. The integer neighborhood of  $\tv'_b$ is comprised of  $27$ integer positions. A fractional refinement of $\tv'_b$ is a sub-voxel offset $\tv_b = (t_x, t_y, t_z) \in [-1,1]^3$ selected within one of these unit cubes. We refer to $\tv_b$ as the FvMV of block $b$, so that the overall motion of the block becomes $\tv'_b + \tv_b$.

To limit the signaling overhead, FvMVs are restricted to have a finite resolution. Given a resolution step size $r$  selected from a finite set:
\begin{equation}
\label{equ:FvME_r}
r \in 
\left\{ 
\frac{1}{2}, \frac{1}{4}, \dots, \frac{1}{2^L} 
\right\}, 
\quad L \in \mathbb{Z}^+,
\end{equation}
the set of admissible FvMVs of a single block is
\begin{equation}
\label{equ:FvME_Mr}
\mathcal{N} = 
\left\{ 
\tv = (t_x, t_y, t_z) 
\;\middle|\; 
t_x,t_y,t_z \in \{0,\pm r, \pm 2r,\dots, \pm 1\} 
\right\},
\end{equation}
which uniformly samples $[-1,1]^3$ with step size $r$ along each axis. The corresponding block-wise fractional motion field belongs to
\begin{equation*}
\mathcal{N}_T = 
\left\{ 
\Tm = (\tv_1, \ldots, \tv_B)^\top \in \mathbb{R}^{B \times 3} 
\;\middle|\; 
\tv_b \in \mathcal{N},\ \forall b 
\right\}.
\end{equation*}

%Because the points are Morton-ordered, point in the same block occupy contiguous rows, so both the target colors \eqref{equ:C_target} and the motion-compensated predictors \eqref{equ:IvMC} can be partitioned block by block. 
The original colors \eqref{equ:C_target} and the integer motion compensated colors \eqref{equ:IvMC} can be organized per block as follows:
\begin{align}
\label{equ:C_target2}
\Cm &=\left[\Cm_1^\top,\;\cdots,\;\Cm_B^\top\right]^\top \in \mathbb{R}^{N\times 3},\\
\label{equ:IvMC2}
\tilde{\Cm}(\Tm)  &= [\tilde{\Cm}_1(\tv'_1)^\top,\hdots,\tilde{\Cm}_B(\tv'_B)^\top]^\top  \in \mathbb{R}^{N\times 3},
\end{align}
where $\Cm_b$ and $ \tilde{\Cm}_b(\tv'_b)\in \mathbb{R}^{N_b\times 3}$.

Since the reference frame contains no attribute samples at fractional positions, we define the FvMC predictor at a fractional displacement as a linear combination of the IvMC predictors over the local $3\times3\times3$ neighborhood. With $\tv_{(1)},\ldots,\tv_{(27)}$ in $\mathcal{M}$ defined in \eqref{equ:M}, the IvMCs at the $27$ integer positions surrounding its motion position are collected to form FvMC predictors of $b$-th block $\hat{\Cm}_{b,27}\in \mathbb{R}^{3N_b\times 27}$:
\begin{equation}
\label{equ:FvMC_IvMC}
\hat{\Cm}_{b,27} = \big(\vec(\tilde{\Cm}_b(\tv'_b+\tv_{(1)})), \ldots,\vec(\tilde{\Cm}_b(\tv'_b+\tv_{(27)}))\big),
\end{equation}
whose $i$-th column is the vectorized IvMC obtained by displacing block $b$ to $\tv'_b+\tv_{(i)}$.

To map a FvMV to a FvMC, we attach to each admissible fractional displacement a sparse trilinear coefficient vector $\xv(\tv) \in \mathbb{R}^{27}$, whose entries depend on $\tv \in \mathcal{N}$, and collect the block-wise coefficients into
\begin{equation}
\label{equ:set_X}
\mathcal{X} = \left\{ \Xm(\Tm) = (\xv(\tv_1),\hdots,\xv(\tv_B)) \in \mathbb{R}^{27\times B} \;\middle|\; \Tm\in\mathcal{N}_T \right\}.
\end{equation}
The FvMC predictor $\hat{\Cm}_{b}(\tv) \in \mathbb{R}^{N_b\times 3}$ associated with a fractional displacement $\tv \in \mathcal{N}$ is then the corresponding combination of the IvMC candidates:
\begin{equation}
\label{equ:FvMC_obj}
\vec(\hat{\Cm}_{b}(\tv)) = \hat{\Cm}_{b,27}\, \xv(\tv).
\end{equation}

FvME selects the fractional motion field that minimizes the energy of the MC residual, accumulated over all points and all three color channels, as a proxy for its coding rate:
\begin{equation}
\label{equ:FvME_obj}
\hat{\Tm}_{\text{F}} = \argmin_{\Tm=(\tv_1, \ldots, \tv_B)^\top \in \mathcal{N}_T} \sum_{b=1}^{B} \left\| \Cm_b - \hat{\Cm}_{b}(\tv_b) \right\|_F^2,
\end{equation}
which is the fractional-precision counterpart of the integer-voxel criterion \eqref{equ:m1}.
Since the predictor of every point depends only on the FvMV of its own block, the objective in \eqref{equ:FvME_obj} separates across blocks, and the search decouples into $B$ independent problems with $\left(2/r+1\right)^3$ candidate displacements each. Exhaustive enumeration is feasible at coarse resolutions, but its cost grows cubically in $1/r$ as the precision is refined, e.g., from $729$ candidates per block at $r = 1/4$ to $4{,}913$ at $r = 1/8$.
Moreover, \eqref{equ:FvME_obj} can be recast as a quadratic program over the coefficient matrices, i.e., over $\mathcal{X}$ and its continuous counterpart. However, neither the finite set $\mathcal{X}$ nor its continuous counterpart is convex, so the problem cannot be handled directly with standard convex optimization tools. 
Based on these observations, we propose a two-stage strategy in which a continuous solution is first found (independent of $r$) and then quantized to resolution $r$.
 
\subsection{FvME modeling}
\label{ssec:FvME_opt}
\subsubsection{Convex relaxation}
Expanding the squared error in \eqref{equ:FvME_obj} and discarding the term $\sum_{b=1}^{B}\|\vec(\Cm_b)\|_2^2$ that does not depend on $\Xm$, the objective becomes a quadratic on the interpolation coefficients:
\begin{equation}
\label{equ:FvME_quad}
F(\Xm) = \sum_{b=1}^{B} F_b(\xv_b),
\end{equation}
where $F_b(\xv_b) = \xv_b^\top \Qm_b$, $\Qm_b = \hat{\Cm}_{b,27}^\top\hat{\Cm}_{b,27}$ and $\qv_b = \hat{\Cm}_{b,27}^\top\vec(\Cm_b)$.  The optimization in
 \eqref{equ:FvME_obj} is equivalent to minimizing $F(\Xm)$ over the set $\mathcal{X}$ of \eqref{equ:set_X}:
\begin{equation}
\label{equ:FvME_obj2}
\hat{\Xm}_{\text{F}} = \argmin_{\Xm=(\xv_1, \ldots, \xv_B) \in \mathcal{X}} \sum_{b=1}^{B} F_b(\xv_b).
\end{equation}
Each $\Qm_b$ is positive semi-definite, so every $F_b$, and therefore $F$, is convex. The difficulty in solving \eqref{equ:FvME_obj2} is not the objective but the feasible set, $\mathcal{X}$, which is discrete. We relax this problem by optimizing over the set
\begin{equation}
\label{equ:FvME_D}
\mathcal{D} = \left\{ \Xm=(\xv_1, \ldots, \xv_B) \in \mathbb{R}^{27\times B} \;\middle|\; \xv_b \in \mathcal{E},\ \forall b \right\},
\end{equation}
where $\mathcal{E} = \{ \xv \in \mathbb{R}^{27} \mid \xv \ge \mathbf{0},\ \mathbf{1}^\top\xv = 1 \}$ is the $27$-dimensional probability simplex. Since every trilinear coefficient vector $\xv(\tv)$ is a convex combination of the $27$ vertices with at most eight nonzero entries, $\mathcal{X} \subset \mathcal{D}$.
Replacing $\mathcal{X}$ by $\mathcal{D}$ in \eqref{equ:FvME_obj2} yields the relaxed problem
\begin{equation}
\label{equ:FvME_obj3}
\hat{\Xm}^* = \argmin_{\Xm=(\xv_1, \ldots, \xv_B) \in \mathcal{D}} \sum_{b=1}^{B} F_b(\xv_b),
\end{equation}
which is a  convex quadratic program that  can be solved to global optimality with standard techniques. Moreover, since the objective is separable, \eqref{equ:FvME_obj3} decouples into $B$ independent $27$-dimensional subproblems.

% \subsubsection{Frank--Wolfe optimization}
% Among such techniques, we adopt the Frank--Wolfe algorithm, applied to each block independently. FW is particularly well suited to these simplex-constrained problems. It is projection-free, its linear subproblem amounts to locating the smallest entry of a $27$-dimensional gradient, and the line search admits a closed form for quadratic objectives. These lightweight iterations are the main source of the computational efficiency of the proposed FvME.

% The iterates are initialized at the simplex vertex corresponding to the integer motion position $(0,0,0)$,
% \begin{equation}
% \label{equ:FvME_initial}
% \xv_{b,0} = \ev_{(0,0,0)} \in \mathcal{E},
% \end{equation} 
% where $\ev_{(0,0,0)}$ is the canonical basis vector at the index of the IvMC offset $\tv = (0,0,0)$. This one-hot start biases the iteration toward small sub-voxel displacements---the typical regime for fractional refinement---and combines with the cube-locking mechanism below to drive the iterate's support onto the eight corners of a single unit cube, matching the sparsity of the discrete trilinear vectors in $\mathcal{X}$.
 
\subsubsection{Frank-Wolfe optimization}
We adopt the Frank-Wolfe algorithm \cite{FW}, applied to each block independently, which is particularly well-suited for simplex-constrained problems: it is projection-free, its linear subproblem amounts to locating the smallest entry in a $27$-dimensional gradient, and the line search admits a closed-form expression for quadratic objectives. These lightweight iterations are the main source of the computational efficiency of the proposed FvME.

\begin{figure}
    \centering
    \includegraphics[width=1\linewidth]{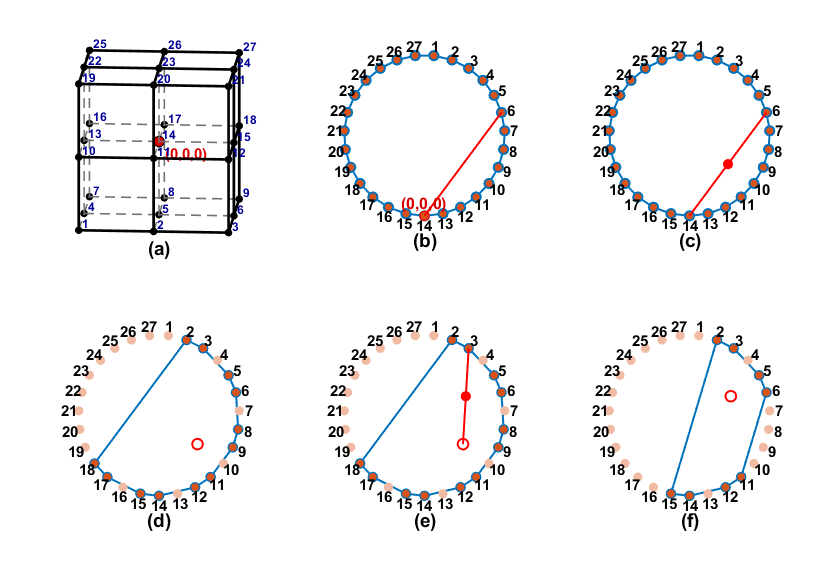}
    \caption{Steps of the proposed FvME for one block. (a) Geometric view: the integer MV $\mathbf{t}'_b$ (red) sits at the center of its $3\times3\times3$ neighborhood, whose 27 IvMC candidates form eight unit cubes. (b)--(f) Frank--Wolfe on the convex hull of IvMCs of these displacement candidates (iterate marked by a red open circle): starting from the center vertex, linear minimization successively selects vertices $14\rightarrow6\rightarrow3$ that belong to the vertices of the bottom right cube. (b) linear minimization; (c) line search; (d) cube locking restricts the active set to a subset of original $27$ IvMCs;  (e) further iterations refine the iterate within that the subset; (f) cube locking finally restrict IvMCs of eight displacement of the right bottom cube, after which the relaxed solution (the red open circle) is associated to the final FvMV.}
    \label{fig:integer_8cubes}
\end{figure}

The $27$ IvMC candidates of a block form a $3\times3\times3$ grid in which eight unit cubes meet at the center offset $\tv_{(14)}=(0,0,0)$ (\autoref{fig:integer_8cubes}). Every admissible FvMV lies in exactly one of these cubes, so its trilinear coefficient vector $\xv(\tv)$ is supported on that cube's eight corners. 
We exploit this structure by initializing the iteration at the center vertex, which is shared by all eight cubes,
\begin{equation}
\label{equ:FvME_initial}
\xv_{b,0}=\ev_{(14)} \in \mathcal{E},
\end{equation}
where $\ev_{(14)}$ is the one-hot vector selecting the IvMC at $\tv_{(14)}$ in \eqref{equ:FvMC_IvMC}, and by progressively confining the search to a single cube, once it is identified (see \emph{Cube locking} below). For block $b$, each $k$-th iteration consists of three steps:
\begin{enumerate}
\item \textbf{Gradient computation.}
Given the current iterate $\xv_{b,k-1}$, the gradient of the $b$th quadratic objective is
\begin{equation}
\label{equ:FvME_grad}
\nabla F_b(\xv_{b,k-1}) = 2\left(\Qm_b\,\xv_{b,k-1}-\qv_b\right) \in \mathbb{R}^{27}.
\end{equation}
\item \textbf{Linear minimization.} The Frank--Wolfe vertex is selected from the current active vertex set $\mathcal{A}_{b,k}\subseteq\{1,\ldots,27\}$ (defined under \emph{Cube locking}):
\begin{equation}
\label{equ:FvME_1hot}
\xv'_{b,k} = \argmin_{\xv\in\,\mathrm{conv}\{\ev_i:\,i\in\mathcal{A}_{b,k}\}} \langle\nabla F_b(\xv_{b,k-1}),\xv\rangle.
\end{equation}
A linear objective over this simplex face attains its minimum at a vertex, so $\xv'_{b,k}=\ev_{i^*}$ with $i^*=\argmin_{i\in\mathcal{A}_{b,k}}[\nabla F_b(\xv_{b,k-1})]_i$, i.e., the index of the smallest gradient entry among the active vertices.
\item \textbf{Line search.} With $\dv_{b,k} = \xv'_{b,k} - \xv_{b,k-1}$, and step size
\begin{equation}
\gamma^*_{b,k} = \argmin_{\gamma\in[0,1]}  F_b\left((1-\gamma)\,\xv_{b,k-1}+\gamma\,\xv'_{b,k}\right)
\end{equation}
can be written in closed form since $F_b$ is quadratic, so the update is
\begin{equation}
\label{equ:FvME_sol}
\xv_{b,k} = (1-\gamma^*_{b,k})\,\xv_{b,k-1}+ \gamma^*_{b,k}\,\xv'_{b,k}.
\end{equation}
\end{enumerate}

\textbf{Cube locking.} Since FW starts at the shared center vertex $\tv_{(14)}$ and refines the predictor by repeatedly selecting one of the $27$ vertices, each selected vertex reveals the corner direction, and hence the cube, toward which the fractional motion is heading. We use this to shrink the active set $\mathcal{A}_{b,k}$: initially $\mathcal{A}_{b,0}=\{1,\ldots,27\}$, and once the selected vertices have determined the sign along all three axes, the enclosing cube is fixed and $\mathcal{A}_{b,k}$ is reduced to the corresponding eight corners, used for the remaining iterations (see  \autoref{fig:integer_8cubes}). 

Each block runs its iterations until $\gamma^*_{b,k}<\epsilon_\gamma$, at which point the block is considered converged and locked to its current iterate $\xv_b^*$. All blocks terminate after at most $f_{\max}$ iterations.
 
\subsubsection{From relaxed weights to the FvMV}
\label{ssec:FvMV_quant}
Geometrically, the relaxation enlarges the feasible set of predictors from trilinear interpolations to arbitrary convex combinations of the $27$ IvMC candidates of each point, i.e., the predicted color of each point is allowed to lie anywhere within the convex hull of its $27$ IvMC candidates in $\mathbb{R}^3$. The relaxed solution $\Xm^*$ generally does not coincide with any valid trilinear coefficient matrix, i.e., $\Xm^* \notin \mathcal{X}$ in general. It is therefore quantized to the nearest matrix in $\mathcal{X}$, and the FvMV $\hat{\Tm}_{\text{F}}$ is the grid motion field associated with the quantized coefficients:
\begin{equation}
\label{equ:FvME_mv}
\hat{\Tm}_{\text{F}} = \argmin_{\Tm \in \mathcal{N}_T} \sum_{b=1}^{B} \big\| \xv_b^* - \xv(\tv_b) \big\|_2^2,
\end{equation}
where $\xv_b^*$ is the $b$th column of $\Xm^*$. 
Due to the cube locking update during linear minimization, $\xv_b^*$ is confined to a single cube. Rounding $\xv_b^*$ to \eqref{equ:FvME_Mr} efficiently produces a valid trilinear coefficient vector for a given resolution $r$.

The selected FvMV $\hat{\Tm}_{\text{F}}$ is entropy coded and transmitted to the decoder as motion overhead. Since the trilinear mapping $\Xm(\cdot)$ of \eqref{equ:set_X} is known at both the encoder and the decoder, the interpolation coefficients $\Xm(\hat{\Tm}_{\text{F}})$ are recovered from $\hat{\Tm}_{\text{F}}$ without any additional signaling, and the FvMC predictor is synchronously constructed at the decoder end as \eqref{equ:FvMC_obj}.

% \begin{equation}
% \label{equ:FvME_mc}
% \vec\!\left(\tilde{\Cm}_{\text{F}}^\top\right) = \tilde{\Cm}_{b,27} \xv(\hat{\tv}_{b}).
% \end{equation}

% The complete procedure is summarized in \autoref{alg:ME}.
% \begin{algorithm}
% \caption{Proposed interpolation-free FvME Algorithm}\label{alg:ME}
% \KwData{IvMC candidates $\tilde{\Cm}_{27}$ and target colors $\Cm$}
% \KwResult{FvMV $\hat{\Tm}_{\text{F}}$ and FvMC $\tilde{\Cm}_{\text{F}}$}
% Initialize $\Xm_0$ via \eqref{equ:FvME_initial} and set $k = 1$\;
% \While{not all blocks converged and $k \leq k_{\max}$}{
% Compute the gradient $\nabla F(\Xm_{k-1})$ via \eqref{equ:FvME_grad}\;
% Find the one-hot vertex matrix $\Xm'_k$ via \eqref{equ:FvME_1hot}\;
% Compute the step size $\gamma^*_{b,k}$ of each block via \eqref{equ:FvME_gamma}\;
% Update $\Xm_k$ via \eqref{equ:FvME_sol} and set $k \leftarrow k+1$\;
% }
% Set $\Xm^*$ to the last iterate\;
% Quantize $\Xm^*$ to obtain the FvMV $\hat{\Tm}_{\text{F}}$ via \eqref{equ:FvME_mv}\;
% Compute the interpolation coefficients $\Xm(\hat{\Tm}_{\text{F}})$ via \eqref{equ:FvME_gridcoeff}\;
% Construct the FvMC $\tilde{\Cm}_{\text{F}}$ via \eqref{equ:FvME_mc}\;
% \end{algorithm}

\section{Experiments}
\label{sec:experiments}
% \subsection{Experimental setting}
\subsection{Dataset and parameter settings}
We evaluate the proposed motion-based inter-coding scheme for the compression of color attributes on the MPEG point cloud sequences \texttt{longdress}, \texttt{redandblack}, \texttt{loot}, \texttt{soldier}, and \texttt{queen}, at a depth resolution of $10$. We use all frames from each sequence for the full test. There are 250 frames for \texttt{queen} and 300 frames for the others. 

We implement our proposed DPC attribute inter-coding system in a low-delay P configuration, using previously decoded frames as references. In a group of frames (GoF), containing $32$ frames, the first frame is coded in intra mode via RAHT from G-PCC \cite{IRAHT2019}, while all blocks in the remaining frames are coded in inter mode using our proposed scheme.  MC residues are also transformed by RAHT. 
The quantized RAHT coefficients and overheads are entropy-coded by the arithmetic coder. 
To avoid high overheads, we use block sizes of $16$, $32$, and $64$ for the high-, mid-, and low-rate regions, respectively. For each rate point, the block size is uniform within and across frames.
%
%
% To reduce the complexity of nearest point search in \autoref{ssec:coarse_search} we match each predicted block to a larger region (a block of size $61\times61\times61$) centered around the zero motion point in the reference frame. 
For graph-based IvME, we choose $\beta_p=0.3$ in \eqref{equ:NNcriterion}, $\beta=10$ in \eqref{equ:rough_model}, $k_{\max}=15$ and $l_{\max}=1$ for optimization \autoref{sssec:optimization}. For FvME,  we choose $f_{\max}=4$ for Frank-Wolfe iterations and $r=\frac{1}{4}$ motion precision for FvMV quantization.

% The overheads $m_{t,b}$s, $k_{t,b}$s and $\hv$ are separately entropy coded.

%\subsection{Evaluation Metrics}
%
% Among the comparison with geometry-based schemes, decoded quality is measured in average peak signal-to-noise ratio over Y component (PSNR-Y),
% \begin{equation}
% \text{PSNR-Y} = \sum_{t=1}^{T} -\frac{10}{T} \log_{10} \left ( \frac{ \| \Ym_t - \Ym'_t  \|_2^2 }{ 255^2 \times N_t} \right), 
% \label{equ:psnrY}
% \end{equation}
% where $\Ym_t$ and $\Ym'_t$ represent original and reconstructed luma signals of the $t$-th frame, respectively, $N_t$ is the total number of input points in $t$-th frame, and $T$ is the total number of frames. 
Since HEVC encodes projected images in YUV420 format, when comparing with V-PCC, decoded quality is measured in the RGB domain to account for chroma sampling. For a single frame, the PRSN is 
\begin{equation}
-10\log_{10} \left ( \frac{ \| \Rm_t - \Rm'_t \|_2^2 + \| \Gm_t - \Gm'_t \|_2^2 + \| \Bm_t - \Bm'_t \|_2^2 }{ 3 \times 255^2  \times N_t} \right).
\label{equ:psnrRGB}
\end{equation}
The PSNR of a point cloud sequence is the average of frame PSNRs. 
Rate is measured in bits per input point (bpip),
\begin{equation*}
\textnormal{bpip} = \frac{\sum_{t = 1}^T b_t}{\sum_{t = 1}^T N_t},
\label{equ:bpip}
\end{equation*}
$b_t$ is the total bitrate required to encode YUV components and motion overhead (when necessary) of the $t$-th frame.  $N_t$ is the total number of input voxels in the $t$-th frame. $b_t$ does not include geometry or occupancy bitrate.

\subsection{Lossless Geometry and Lossy Attribute Case}

\begin{table}[H]
  \centering
\begin{tabular}{ |p{1.4cm}||p{0.3cm}| p{2cm} | p{0.7cm} | p{1.9cm}|}
% \begin{tabular}{ |c||c|c|c|c|}
\hline
Test Schemes & GOF & Lossless geometry & Recolor& Lossy attribute  \\
 \hline
Proposed & 32 &  Octree & N/A & Proposed + RAHT \\
\hline
V-PCCv24 & 32 & HEVC-LD & N/A & HEVC-LD  \\
\hline
GeS-TMv3 & 32 & Octree & N/A & InterRAHT  \\
\hline
G-PCCv23 & 1 & Octree & N/A & RAHT \\
 \hline
\end{tabular}
 \caption{Settings summary of our proposed approach and anchor coding solutions in the experiments on \textbf{lossless} geometry condition. Since geometry is lossless, color transfer is not needed.} 
 \label{tab:experiment1_setting}
% \vspace{-8mm}
\end{table}

\begin{figure*}[htb]
\begin{center}
\includegraphics[width = 1\linewidth]{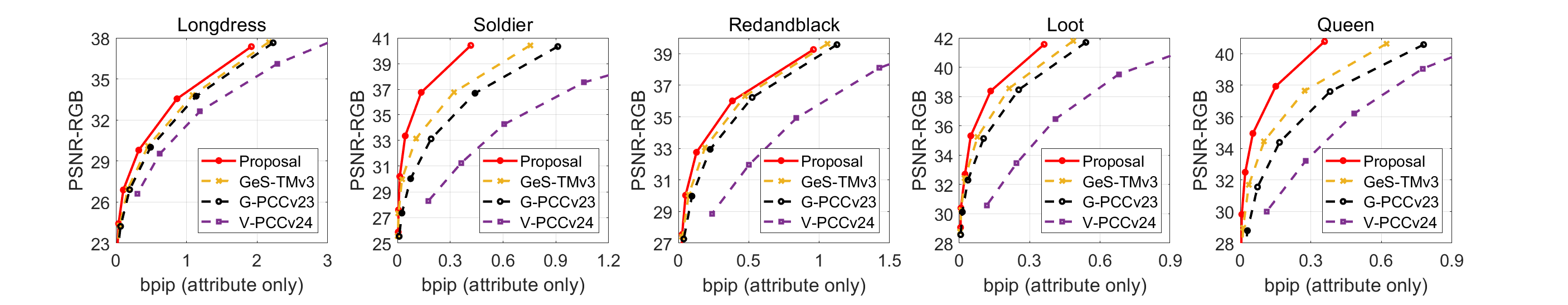}
\caption{Comparison with the baselines on the \autoref{tab:experiment1_setting} settings}
\label{fig:lossless}
\end{center}
\end{figure*}

We apply the \textit{lossless geometry coding} setting to G-PCCv23 \cite{GPCC}, GeS-TMv3 \cite{sandri2023}, and V-PCCv24 \cite{VPCC}. For G-PCC and GeS-TM, we use octree-based lossless geometry coding. For V-PCC, both raw and EOM patches are enabled to achieve lossless geometry encoding. Results are shown in \autoref{fig:lossless}. The proposed method achieves BD-rate reductions of $-51.3\%$, $-31.7\%$, and $-146.4\%$ compared to G-PCC,  GeS-TM, and V-PCC, respectively.

Compared with GeS-TM, our proposed scheme adopts a more advanced ME scheme tailored to color, using color distance, motion coherence, and higher motion precision during search, resulting in more accurate predictors. 
% Our motion compensation filtering scheme can refine MC to reduce coding rates and improve reconstruction quality. 
The proposed scheme outperforms G-PCC by further reducing temporal redundancy, while G-PCC can only remove spatial redundancy.
Our proposed scheme achieves significant coding gain at all rates compared to V-PCC. 
V-PCC cannot remove the high-frequency redundancy among the sequentially placed voxels in the raw and EOM texture patches, which appear as noise-like patterns, resulting in the worst performance under the lossless-geometry coding setting.

\begin{table}[H]
  \centering
\begin{tabular}{ |c||c|c|c|}
\hline
\multirow{2}{*}{Point Clouds}
 & \multicolumn{3}{|c|}{BD-BR (\%)}  \\\cline{2-4}
 &  Proposed & GeS-TMv3 & V-PCCv24  \\
 \hline
Longdress   & \textbf{-30.3} & -11.4 & 56.1 \\
\hline
Soldier     & \textbf{-74.7} & -51.9 & 192.3 \\
\hline
Redandblack & \textbf{-30.7} & -20.5 & 151.6 \\
\hline
Loot        & \textbf{-46.5} & -28.4 & 243.4 \\
\hline
Queen       & \textbf{-74.4} & -46.5 & 88.9 \\
\hline
\hline
Average     & \textbf{-51.3} & -31.7 & 146.4 \\
\hline
\end{tabular}
 \caption{Attribute BD-BR of our proposed approach compared to state-of-the-art baselines on the \autoref{tab:experiment1_setting} settings} 
 \label{tab:results_BDBR_lossless} 
% \vspace{-8mm}
\end{table}

\subsection{Lossy Geometry and Lossy Attribute Case}

\begin{table}[H]
  \centering
\begin{tabular}{ |p{1.4cm}||p{0.3cm}| p{1.8cm} | p{1.0cm} | p{1.8cm}|}
\hline
Test Schemes & GOF & Lossy geometry & Recolor& Lossy attribute \\
 \hline
Proposed & 32 &  HEVC-LD & V-PCC & Proposed + RAHT \\
\hline
V-PCCv24 & 32 & HEVC-LD & V-PCC & HEVC-LD  \\
\hline
GeS-TMv3 & 32 & HEVC-LD & V-PCC & InterRAHT  \\
\hline
G-PCCv23 & 1 & HEVC-LD & V-PCC & RAHT \\
 \hline
\end{tabular}
 \caption{Settings summary of our proposed approach and attribute coding anchors on \textbf{lossy} geometry condition} 
 \label{tab:experiment2_setting}
% \vspace{-8mm}
\end{table}

% \begin{table}[H]
%   \centering
% \begin{tabular}{ |p{1.4cm}||p{0.3cm}| p{1.8cm} | p{1.0cm} | p{1.7cm}|}
% \hline
% Test Schemes & GOF & Geometry (lossy) & Recolor& Attribute (lossy) \\
% \hline
% V-PCCv24 & 1 & HEVC-AI & V-PCC & HEVC-AI  \\
% \hline
% G-PCCv23 & 1 & HEVC-AI & V-PCC & RAHT \\
%  \hline
% \end{tabular}
%  \caption{Experimental settings for comparing attribute coding solutions of G-PCCv23 and V-PCCv24-AI on \textbf{lossy} geometry condition} 
%  \label{tab:experiment3_setting}
%  \vspace{-8mm}
% \end{table}

%This experiment compares the attribute coding schemes after alignment. Specifically, 
We now compare attribute coding schemes \textit{using a common lossy-coded geometry}, obtained via lossy reconstruction from V-PCC under the low-delay configuration, onto which the original color attributes are transferred using the V-PCC recolor function. 
All attribute codecs use the same geometry and recolored attributes, so the comparison isolates attribute-coding efficiency from the geometry coding.
\begin{figure*}[htb]
\begin{center}
    \includegraphics[width = 1\linewidth]{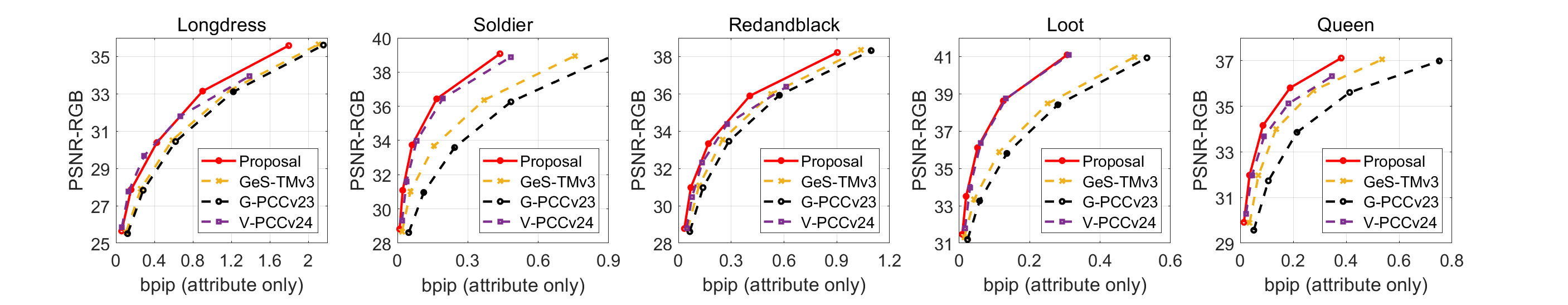}
    \caption{Comparison with the anchors on the \autoref{tab:experiment2_setting} settings}
\label{fig:lossy_vpcc}
\end{center}
\end{figure*}

V-PCC's attribute coding performs much better with lossy geometry than with lossless geometry, because the points assigned to raw patches that are expensive to encode are pruned. Also, the empty pixels between patches are filled with smooth color to avoid drastic color changes across patch boundaries and to reduce high-frequency transform coefficients. As a result, V-PCC's coding efficiency increases significantly and approaches that of our scheme.

Our scheme outperforms V-PCC, achieving $16.5\%$ BD-BR savings on average (\autoref{tab:result_BDBR_lossy}), suggesting that working directly in 3D is beneficial. 
In V-PCC, the generated 2D patches cannot be guaranteed to remain temporally consistent, making accurate motion estimation more difficult and resulting in higher energy in the MC residuals. Additionally, because segmentation for projection breaks the original 3D correlation, more MV signaling is typically needed. 

\begin{table}[H]
  \centering
\begin{tabular}{ |c||c|c|c|}
\hline
\multirow{2}{*}{Point Clouds}
 & \multicolumn{3}{|c|}{BD-BR (\%)}  \\\cline{2-4}
&  Proposed & GeS-TMv3 & V-PCCv24  \\
 \hline
Longdress   & \textbf{-34.7} & -8.1  & -34.5  \\
\hline
Soldier     & \textbf{-74.1} & -39.5 & -66.4  \\
\hline
Redandblack & \textbf{-38.2} & -14.7 1& -22.2  \\
\hline
Queen       & \textbf{-63.1} &-20.9 & -56.2 \\
 \hline
Loot        & \textbf{-66.2} & -38.2 & -55.4 \\
 \hline
 \hline
Average     & \textbf{-55.2} & -24.27 & -46.9 \\
 \hline
\end{tabular}
 \caption{Attribute BD-BR of our proposed approach compared to state-of-the-art anchors on the \autoref{tab:experiment2_setting} settings} 
 \label{tab:result_BDBR_lossy}
 \vspace{-8mm}
\end{table}

\subsection{Complexity Analysis}
\begin{table}[H]
  \centering
% \begin{tabular}{ |p{1.8cm}||p{1.1cm}| p{1.2cm} | p{1.1cm} | p{1.2cm}|}
\begin{tabular}{ |c||c|c|c|c|c|c|}
\hline
\multirow{3}{*}{PCs}
    & \multicolumn{6}{|c|}{Time Complexity (seconds)} \\\cline{2-7}
 & \multicolumn{3}{|c|}{Encoder} & \multicolumn{3}{|c|}{Decoder} \\\cline{2-7}
 & V-PCC &  GeS-TM &  Proposed & V-PCC & GeS-TM &  Proposed \\
 \hline
Seq1 & 2766.4 &  95.6& 603.2 & 42.8 & 24.9& 70.2\\
\hline
Seq2 & 3923.9 & 85.9& 553.4 & 43.1 & 22.2&  62.4\\
\hline
Seq3 & 3413.1 & 126.6& 868.7 & 52.1 & 32.4&  110.7\\
\hline
Seq4 & 3192.8 & 113.1& 760.2 & 58.4 & 27.9& 92.7\\
 \hline
 Seq5 & 4255.6 & 95.6& 668.4 & 39.4 & 24.5& 78.1\\
 \hline
  \hline
  Average & 3550.4 & 103.4& 690.8 & 47.2 & 26.4& 82.8\\
   \hline
Ratio  & \multicolumn{3}{|c|}{34.5:1:6.7} & \multicolumn{3}{|c|}{1.8:1:3.1} \\\cline{2-7}
 \hline

\end{tabular}
 \caption{Encoding and decoding time of a single GoF including 1 I-frame and 31 P-frames. Seq1-Seq5 are Loot, Soldier, Redandblack, Queen, and Longdress.} 
 \label{tab:result_complexity}
% \vspace{-8mm}
\end{table}

% Due to our MATLAB implementation, a direct time comparison with the C++ V-PCC implementation is not possible. 
% We compare the time complexities of our proposed scheme, GeS-TM and V-PCC. 
% Due to our MATLAB implementation, a direct time comparison with the C++ V-PCC implementation is not fair. 
The algorithm runtime reported in \autoref{tab:result_complexity} includes the encoding/decoding of both geometry and attributes. GeS-TM and V-PCC are fully implemented in C++. Our proposed ME is implemented in MATLAB, while we still rely on the C++ implementation GPCC for RAHT, quantization and entropy coding.  
On average, the proposed scheme is slightly slower than GeS-TM but more efficient than V-PCC.
% GeS-TM is $35$ times faster than V-PCC.
In our scheme, the most time-consuming parts are the nearest-point search in motion estimation and motion compensation.
% , and the graph construction for two-pass filtering. 
% All of their complexity can be significantly reduced to $O(NlogN)$ by reusing the KD-tree built during recoloring.
Though our scheme is more complex than GeS-TM due to more complicated motion estimation and compensation, it remains in the same order of magnitude because it is also geometry-based and 3D-native, and it avoids the computational inefficiencies of 2D-to-3D conversion and highly complex 2D video coding.

\section{Conclusion}
We proposed a novel geometry-based motion estimation scheme for dynamic solid point cloud attributes. 
%We propose a complete motion estimation scheme to remove temporal color redundancy. 
The whole coding scheme is implemented on top of the state-of-the-art codec G-PCC. The coding results demonstrate that our proposed method outperforms the widely used G-PCC, GeS-TM, and V-PCC. 
While the proposed ME scheme successfully captures the dominant motion patterns in dynamic point clouds, certain sequences exhibit complex, highly non-rigid motion—such as hair dynamics, garment fluttering, and viscoelastic deformation of soft tissues. As translation motion cannot accurately describe most of these motions, residual energy remains high after motion compensation. Addressing this problem is left for future work.

% $$ \int \mathbf{S} (u-f)^{2} \, dA = \int \mathbf{S} {\sum}{i=1}^n {\sum}{j=1}^n {\varphi}_i\cdot {\varphi}_j (u_i-f_i) (u_j-f_j) \, dA = (\mathbf{u}-\mathbf{f})^{\mathsf T} \mathbf{M} (\mathbf{u}-\mathbf{f}), $$

% $${\chi}_\mathbf{S}(\mathbf{x}) = \begin{cases} 1 & \text{ if $\mathbf{x} \in \mathbf{S}$ } \\ 0 & \text{ otherwise}. \end{cases} $$

% $$
% {\nabla}g(\mathbf{x}_{i,j,k}) = \mathbf{v}_{i,j,k} := \begin{cases}
% \vphantom{\left(\begin{array}{c}
% 0\\
% 0\\
% 0\end{array}\right)}
% \mathbf{n}_\ell & \text{ if $\exists\ \mathbf{p}_\ell = \mathbf{x}_{i,j,k}$}, \\
% \left(\begin{array}{c}
% 0\\
% 0\\
% 0\end{array}\right) & \text{ otherwise}.
% \end{cases}
% $$

\bibliographystyle{ieeetr}
\bibliography{refs.bib}

% \newpage
% \afterpage{\clearpage}
% \appendix
% \subsection{Additional Results}
% \begin{center}
%     \includegraphics[width=\linewidth]{./figures/ablation_IMV.png}
%     \captionof{figure}{Comparison between graph-based and ColorICP-based IvME.}
%     \label{fig:ablation_IVME}
% \end{center}

% \begin{center}
%     \includegraphics[width=\linewidth]{./figures/ablation_FMV.png}
%     \captionof{figure}{Comparison between interpolation-free and interpolation-based FvME.}
%     \label{fig:ablation_FVME}
% \end{center}

% \begin{center}
%     \includegraphics[width=\linewidth]{./figures/ablation_FMV2.png}
%     \captionof{figure}{Comparison on $\frac{1}{2}$ and $\frac{1}{4}$ interpolation-free FvME.}
%     \label{fig:ablation_FVME2}
% \end{center}

\end{document}

%% file: macros.tex
\usepackage{bm}

\newfont{\bbb}{msbm10 scaled 700}

\newcommand{\cv}{{\bf c}}
\newcommand{\dv}{{\bf d}}
\newcommand{\ev}{{\bf e}}
\newcommand{\fv}{{\bf f}}

\newcommand{\mv}{{\bf m}}

\newcommand{\pv}{{\bf p}}
\newcommand{\qv}{{\bf q}}

\newcommand{\tv}{{\bf t}}

\newcommand{\xv}{{\bf x}}

\newcommand{\Bm}{{\bf B}}
\newcommand{\Cm}{{\bf C}}
\newcommand{\Dm}{{\bf D}}

\newcommand{\Gm}{{\bf G}}

\newcommand{\Km}{{\bf K}}
\newcommand{\Lm}{{\bf L}}
\newcommand{\Mm}{{\bf M}}

\newcommand{\Pm}{{\bf P}}
\newcommand{\Qm}{{\bf Q}}
\newcommand{\Rm}{{\bf R}}
\newcommand{\Sm}{{\bf S}}
\newcommand{\Tm}{{\bf T}}
\newcommand{\Um}{{\bf U}}
\newcommand{\Wm}{{\bf W}}
\newcommand{\Vm}{{\bf V}}
\newcommand{\Xm}{{\bf X}}

\newcommand{\Lambdam}{\hbox{\boldmath$\Lambda$}}

\newcommand{\Sigmam}{\hbox{\boldmath$\Sigma$}}

\newcommand{\Omegam}{\hbox{\boldmath$\Omega$}}

\renewcommand{\det}{{\hbox{det}}}

\renewcommand{\arg}{{\hbox{arg}}}

\newcommand{\transp}{{\sf T}}
\renewcommand{\vec}{{\rm vec}}